\documentclass{elsart}
\usepackage{graphicx}% Include figure files
\usepackage{bm}% bold math
\usepackage{amsmath}
\usepackage{amssymb}
\renewcommand{\theequation}{\arabic{equation}}
\numberwithin{equation}{section}

\def\Journal#1#2#3#4{{#1} {#2} (#4) #3 }

\def\NPA{{\em Nucl. Phys.} A}

\def\PRO{{\em Prog. Theor. Phys.}}

\def\PLB{{\em Phys. Lett.} B}

\def\PRL{\em Phys. Rev. Lett.}

\def\PRC{{\em Phys. Rev.} C}

\def\PTP{\em Prog. Theor. Phys.}

\def\ZPA{{\em Z. Phys.} A}

\def\PPNP{{\em Prog. Part. Nucl. Phys.}}

\newcommand{\be}{\begin{equation}}
\newcommand{\ee}{\end{equation}}
\newcommand{\bea}{\begin{eqnarray}}
\newcommand{\eea}{\end{eqnarray}}

\def\mib#1{\hbox{\boldmath $#1$}}
\def\CG{{\mathcal G}}

\def\CP{{\mathcal P}}

\def\CV{{\mathcal V}}

\def\CP{{\mathcal P}}

\def\CY{{\mathcal Y}}

\def\bK{\mbox{\boldmath $K$}}
\def\bL{\mbox{\boldmath $L$}}

\def\bS{\mbox{\boldmath $S$}}
\def\bX{\mbox{\boldmath $X$}}
\def\bk{\mbox{\boldmath $k$}}
\def\bn{\mbox{\boldmath $n$}}
\def\bp{\mbox{\boldmath $p$}}
\def\bq{\mbox{\boldmath $q$}}
\def\br{\mbox{\boldmath $r$}}

\def\bx{\mbox{\boldmath $x$}}

\def\bfell{\mbox{\boldmath $\ell$}}
\def\bfsigma{\mbox{\boldmath $\sigma$}}
\def\bftau{\mbox{\boldmath $\tau$}}
\def\bfxi{\mbox{\boldmath $\xi$}}

\def\eq#1{Eq.\,(\ref{#1})}
\begin{document}
\begin{frontmatter}

%\preprintnumber{%To give preprint# at top right corner when [preprint]
%\thispagestyle{myheadings}
%\markright{
%\hfill KUNS-1624 \\
% November 1999
%\hfill nucl-th/9912047

\title{\ \mib{\Lambda \alpha}, \mib{\Sigma \alpha}
\bf{and} \mib{\Xi \alpha}
\bf{potentials derived from the} \mib{SU_6} {\bf quark-model
baryon-baryon interaction}}

\author{Y. Fujiwara, M. Kohno$^{*}$ and Y. Suzuki$^{**}$
}
%\qquad   \\
\address{
Department of Physics, Kyoto University,
Kyoto 606-8502, Japan \\
$\hbox{}^{*}$Physics Division, Kyushu Dental College,
Kitakyushu 803-8580, Japan \\
$\hbox{}^{**}$Department of Physics, and Graduate School
of Science and Technology, \\
Niigata University, Niigata 950-2181, Japan}

\maketitle

\begin{abstract}
\indent
We calculate $\Lambda \alpha$, $\Sigma \alpha$ and $\Xi \alpha$
potentials from the nuclear-matter $G$-matrices of the $SU_6$ quark-model
baryon-baryon interaction. The $\alpha$-cluster wave function is
assumed to be a simple harmonic-oscillator shell-model
wave function. A new method is proposed to derive the direct and knock-on 
terms of the interaction Born kernel from the hyperon-nucleon $G$-matrices,
with explicit treatments of the nonlocality and
the center-of-mass motion between the hyperon and $\alpha$.
We find that the $SU_6$ quark-model baryon-baryon interactions,
FSS and fss2, yield a reasonable bound-state energy
for $\hbox{}^5_\Lambda \hbox{He}$, $-3.18 \sim -3.62$ MeV,
in spite of the fact that they give relatively large depths
for the $\Lambda$ single-particle potentials,
46 $\sim$ 48 MeV, in symmetric nuclear matter.
An equivalent local potential derived from the Wigner transform
of the nonlocal $\Lambda \alpha$ kernel shows a strong
energy dependence for the incident $\Lambda$-particle, indicating
the importance of the strangeness-exchange process in the
original hyperon-nucleon interaction.  
For the $\Sigma \alpha$ and $\Xi \alpha$ potentials, we only
discuss the zero-momentum Wigner transform of the interaction
kernels, since these interactions turn out to be repulsive when the 
two isospin contributions for the $\Sigma N$ and $\Xi N$
interactions are added up. These components show a strong isospin
dependence: They are attractive in the isospin $I=1/2$ ($\Sigma \alpha$)
and $I=0$ ($\Xi \alpha$) components and repulsive in $I=3/2$
($\Sigma \alpha$) and  $I=1$ ($\Xi \alpha$) components,
which indicate that $\Sigma$ and $\Xi$ potentials
could be attractive in some particular systems
such as the well-known $\hbox{}^4_\Sigma \hbox{He}$ system. 
\end{abstract}

\end{frontmatter}

%\clearpage

\section{Introduction}

Interactions between the octet baryons ($B_8=N,~\Lambda,~\Sigma$ and
$\Xi$) and the $\alpha$ cluster are important ingredients
to consider the possible existence for various kinds of light hypernuclei
through detailed microscopic cluster-model calculations.
If reliable effective $B_8 N$ interactions were known, one could easily
calculate the $B_8 \alpha$ potentials, using the standard
cluster-model techniques. Unfortunately, this is not the case
except for $B_8=N$, since the bare $B_8 N$ interaction itself
is not well known especially for the $\Sigma$ and $\Xi$ hyperons,
because of the technical difficulties of strangeness experiments.
Even if these bare interactions were eventually known, we further need to
develop the procedure to link the bare interactions and effective
interactions through some effective interaction theory such as the
$G$-matrix formalism.

We have recently developed the QCD-inspired spin-flavor $SU_6$ quark
model for the baryon-baryon interaction \cite{PPNP},
which is a unified model for the full octet-baryons \cite{B8B8},
and have achieved accurate descriptions
of the $NN$ and $YN$ interactions \cite{fss2}.
In particular, the $NN$ interaction of the most
recent model fss2 is accurate enough to compare with
modern realistic meson-exchange models.
These quark-model interactions were used
for the detailed study of few-baryon systems
such as $\hbox{}^3\hbox{H}$ \cite{triton,PANIC02}
and $\hbox{}^3_\Lambda \hbox{H}$ \cite{hypt},
and also of some typical $\Lambda$-hypernuclei,
$\hbox{}^9_\Lambda \hbox{Be}$ \cite{2al,2aljj} and 
$\hbox{}^{\ \,6}_{\Lambda \Lambda}\hbox{He}$ \cite{HE6LL},
through a newly developed three-cluster Faddeev formalism \cite{TRGM,RED}
and $G$-matrix calculations \cite{GMAT,SPLS,KO03}.
We can now use these baryon-baryon interactions to calculate not only
the $\Lambda \alpha$ interaction, but also $\Sigma \alpha$ and
$\Xi \alpha$ interactions, assuming the harmonic-oscillator (h.o.)
shell-model wave function for the $\alpha$-cluster.

There are, in fact, a couple of advanced procedure to
derive the interactions between a single baryon and finite
nuclei based on the density-dependent Hartree-Fock theory
of $G$-matrix interactions \cite{CS72,KNY75,KS83}.
These approaches, however, use the localized $G$-matrix
interaction in the configuration space and the center-of-mass (c.m.)
coordinate system connected to the target nucleus.  
For the $B_8 \alpha$ interactions, the c.m. correction is quite
important. In this paper, we will derive $B_8 \alpha$ Born kernels,
directly starting from the $G$-matrix calculation of the
quark-model baryon-baryon interactions.
We deal with the nonlocality of the $G$-matrix
and the c.m. motion exactly,
using the cluster-model approach, at the expense of the 
self-consistency between the $\alpha$-cluster formation
and the $G$-matrix interaction.
The $G$-matrix is pre-determined by solving the Bethe-Goldstone
equation in symmetric nuclear matter,
and the momentum-dependent single-particle (s.p.) potentials
as well \cite{GMAT}.
The Fermi-momentum is assumed to be the standard
value $k_F=1.35~\hbox{fm}^{-1}$, but the obtained $B_8 \alpha$
interactions depend on this choice rather monotonously, except for 
some special cases such as the $\Lambda \alpha$ $LS$ interaction.
The treatment of the starting-energy dependence 
and the non-Galilean invariant momentum dependence
of the $G$-matrix is explicitly discussed.
We believe that these approximations are good enough
to understand the unknown $\Sigma \alpha$
and $\Xi \alpha$ interactions qualitatively, starting from
the quark-model predictions of various baryon-baryon interactions.
For the $\Lambda \alpha$ interaction, we compare the
predictions by the present approach with some available
phenomenological $\Lambda \alpha$ potentials \cite{2al,HI97},
obtained by various methods.
Another application of the present approach to the $N \alpha$
interaction will be published in a separate paper, since
this system involves an extra nucleon-exchange term.

Since the obtained $B_8 \alpha$ interactions are all nonlocal,
we calculate the Wigner transform from the $B_8 \alpha$
Born kernels. We find that the momentum dependence of the
Wigner transform is very strong, and the procedure to
find effective local potential by solving the transcendental
equation is necessary to obtain a local-potential image
for the $\Lambda \alpha$ interaction.
The $\Sigma \alpha$ and $\Xi \alpha$ interactions are repulsive,
although the isospin $I=1/2$ (for $\Sigma \alpha$) and $I=0$ (for $\Xi \alpha$)
contributions of the $\Sigma N$ and $\Xi N$ interactions
are attractive. As the first step to study realistic $\Sigma \alpha$
and $\Xi \alpha$ interactions, we will only discuss the zero-momentum 
Wigner transform in this paper. Although these $B_8 \alpha$ interactions
are complex due to the imaginary part of the underlying $G$-matrices,
we discuss only the real part. The spin-orbit $B_8 \alpha$ potentials
are naturally obtained from the $LS$ and $LS^{(-)}$ components
of the $G$-matrix invariant interaction.

The central $\Lambda\alpha$, $\Sigma\alpha$ and $\Xi\alpha$
potentials were calculated by many authors with other (usually
more crude or purely phenomenological) approaches. 
The $\Lambda\alpha$ potential was calculated numerously,
and different shapes were considered.
For instance, see Refs.\,\cite{DA82,NA84,GU93,FI03}.
Some phenomenological $\Sigma \alpha$ and $\Xi \alpha$ potentials
are found in Ref.\,\cite{HA90,OK90,YA94,MY94}.
The comparison of our results with these potentials are, however,
not easy, since our Wigner transform is momentum-dependent
and the solutions of the transcendental equations
are strongly energy-dependent.
We will only point out if some resemblance between them
are found. The importance of the isospin dependence
of the $\Sigma N$ and $\Xi N$ interactions
to the $\Sigma$-nucleus and $\Xi$-nucleus potentials 
has been discussed by many authors from the
phenomenological aspects, for example,
in Refs.\,\cite{DO84,HA90,YA94}, and by Rijken and Yamamoto
from the viewpoint of $G$-matrix calculations in Ref.\,\cite{RI06}.

The organization of this paper is as follows.
In the next formulation section, we first give in Section 2.1
the basic folding formula for the $\alpha$-cluster
based on the separation of the $B_8 \alpha$ Born kernel
to the spin-isospin factors and the spatial part.
The treatment of $G$-matrix variables, the starting energy
and the c.m. momentum, will be carefully discussed.
A convenient transformation formula for the rearrangement
of relative momenta is given in Section 2.2, by which
the partial wave component of the $B_8 \alpha$ Born
kernel is explicitly given both for the central component
and the $LS$ component. 
The folding formula in the partial-wave expansion and
the partial-wave components of the $B_8 \alpha$ Born kernel
are explicitly given in Section 2.3. 
A procedure to calculate the
Wigner transform is given in Section 2.4 with a couple of
approximations. One of the approximations for the $LS$ component
yields a simple factor for the strength of the $LS$ potential,
which corresponds to the well-known Scheerbaum
factor $S_B$ \cite{SC76} in nuclear matter. 
The Section 3 is devoted to the results and discussion; first
in Section 3.1 the $\Lambda \alpha$ central and $LS$ potentials
both in the $T$-matrix approach and in the Wigner transform
approach. The isospin dependence of the $\Sigma \alpha$ and
$\Xi \alpha$ potentials is discussed in Section 3.2.
Section 4 is devoted to a summary.
The invariant $G$-matrix for the most general $B_8 B_8$ interaction
is discussed in Appendix A. 

\section{Formulation}

\vspace{-3mm}

\subsection{$\alpha$-cluster folding for the
$G$-matrix invariant interaction}

\vspace{-3mm}

The $B_8 \alpha$ Born kernel for the $(0s)$ h.o. $\alpha$-cluster
wave function is calculated from
\begin{equation}
V(\bq_f, \bq_i)=\langle\,\delta(\bX_G) e^{i \bq_f \cdot \br}
\chi_B \phi_\alpha\,|\,\sum^5_{j=2} G_{1j}\,|\,1 \cdot
e^{i \bq_i \cdot \br} \chi_B \phi_\alpha\,\rangle \ \ ,
\label{fm1-1}
\end{equation}
where $\chi_B$ is the spin-isospin wave function of $B$,
$\phi_\alpha$ the internal wave function
of $\alpha$, and $G_{1j}$ the $BN$ $G$-matrix acting on the
particle $i=1$ ($B$) and $j=2$ - 5 (nucleons).
We use the short-hand notation $B$ to specify one of the
octet baryons, $B_8= N,~\Lambda,~\Sigma$ or $\Xi$.
In \eq{fm1-1}, it is important to calculate the Born kernel
in the total c.m. system by inserting $\delta(\bX_G)$ and
1 in the bra and ket sides, respectively \cite{NA95}, since
the $G$-matrix interaction $G_{1j}$ is non-Galilean invariant.
Namely, the two-particle $G$-matrix which satisfies
the translational invariance is parametrized by
\begin{equation}
\langle \bp_1, \bp_2 | G | \bp^\prime_1, \bp^\prime_2 \rangle
=\delta(\bK-\bK^\prime)\frac{1}{(2\pi)^3}
G(\bp, \bp^\prime; K, \omega, k_F)\ \ .
\label{fm1-2}
\end{equation}
Here, $\omega$ is the starting energy, $K=|\bK|$ is
the magnitude of the c.m. momentum, $k_F$ is the Fermi momentum
of nuclear matter, and the relative momentum $\bp$ (and also
$\bp^\prime$ etc. with primes),
is given by
\begin{eqnarray}
\bp & = & \frac{\xi \bp_1-\bp_2}{1+\xi}\ \ ,\qquad
\qquad \bp_1=\frac{1}{1+\xi}\bK+\bp\ \ ,\nonumber \\
\bK & = & \bp_1+\bp_2\ \ ,\qquad \qquad \bp_2=\frac{\xi}{1+\xi}\bK-\bp\ \ ,
\label{fm1-3}
\end{eqnarray}
with $\xi=(M_N/M_B) \leq 1$ being the mass ratio between the nucleon
and the baryon $B$.\footnote{Note that $\zeta$ used in Appendix B
of Ref.\,\cite{2al} is the inverse of $\xi$.} 
We assume a constant $k_F=1.35~\hbox{fm}^{-1}$ in this paper
unless otherwise specified, so that we will omit this index
in the following.
In fact, the $G$-matrix in \eq{fm1-1} contains the exchange term.
It is, therefore, convenient to use the isospin sum 
of the invariant $G$-matrix, $G^{I}_{BB}(\bp, \bp^\prime; K, \omega)$
with $I=I_B+1/2$ and $I_B-1/2$ ($I_B$ is the isospin of $B$), defined 
through
\begin{eqnarray}
& & G^{I}_{BB}(\bp, \bp^\prime; K, \omega) \nonumber \\
& & =\langle\,[BN]_{II_z}\,|\,G(\bp, \bp^\prime; K, \omega)
- G(\bp, -\bp^\prime; K, \omega)\,P_\sigma\,P_F\,|
\,[BN]_{II_z}\,\rangle\ \ \nonumber \\
& & = \left(1+\delta_{B,N}\right)\left[
g^I_0+g^I_{ss} (\bfsigma_1 \cdot \bfsigma_2)
+h^I_0\,i \widehat{\bn} \cdot (\bfsigma_1 + \bfsigma_2)
+h^I_-\,i \widehat{\bn} \cdot (\bfsigma_1 - \bfsigma_2)
\right. \nonumber \\
& & \hspace{29mm} \left. + \cdots \right]
\label{fm1-4}
\end{eqnarray} 
Here $\widehat{\bn}=[\bp^\prime \times \bp]/(p^\prime p \sin \theta)$,
and the invariant functions $g^I_0$ (central), $g^I_{ss}$ (spin-spin),
$h^I_0$ ($LS$), $h^I_-$ ($LS^{(-)}$), etc. are functions of $\bp^2$,
${\bp^\prime}^2$, $\cos \theta=(\widehat{\bp} \cdot \widehat{\bp}^\prime)$,
$K$, $\omega$, and $k_F$. These are expressed by the partial-wave
components of the $BN$ $G$-matrix as (see Appendix D of Ref.\,\cite{LSRGM})
\begin{eqnarray}
\left. \begin{array}{c}
g^{I}_0 \\
g^{I}_{ss} \\ 
\end{array}
\right \}
& = & \frac{1}{4}\,\sum_{J \ell S} (2J+1)
\left\{ 
\begin{array}{c}
 1 \\ 
\frac{1}{3}[2S(S+1)-3] 
\end{array}
\right\} 
G^{IJ}_{S \ell, S \ell}~P_l(\cos\theta)\ \ , \nonumber \\
h^{I}_0 & = &  - \frac{1}{4} \sum_J 
\frac{(2J+1)}{J(J+1)}
\left[ G^{IJ}_{1 J, 1 J}~P^1_J(\cos\theta) \right.
\nonumber\\
& &  \left. + J~G^{IJ}_{1 J+1, 1 J+1} 
~P^1_{J+1}(\cos\theta)
- (J+1) G^{IJ}_{1 J-1, 1 J-1}
~P^1_{J-1}(\cos\theta) \right]\ \ , \nonumber \\
h^{I}_- & = & \frac{1}{4} \sum_J 
\frac{(2J+1)}{\sqrt{J(J+1)}}
\left[ G^{IJ}_{1 J, 0 J} + G^{IJ}_{0 J, 1 J}
\right]~P^1_J(\cos\theta)\ \ ,
\label{fm1-5}  
\end{eqnarray}
where the argument $\bp$, $\bp^\prime$, $K$, $\omega$, $k_F$
and subscripts for the baryon channels are omitted
for the typographical reason.
In the $LS$ term, $P^1_J(\cos\theta)
=(\sin\theta) P^\prime_J(\cos\theta)$ with $J \geq 1$ is
the associated Legendre function of the first rank.
The invariant $G$-matrix for the most general $B_8 B_8$ interaction
is discussed in Appendix A. 

In order to calculate the spin-isospin factors, it is convenient
to introduce an isospin $pseudo$-exchange operator of the $B N$ system
by
\begin{equation}
P_\tau=\frac{1}{2I_B+1}\left(1+\bftau_B \cdot \bftau_N\right)\ \ ,
\label{fm1-6}
\end{equation}
with the isospin matrix elements
\begin{equation}
\left(\bftau_B \cdot \bftau_N\right)
=\left\{\begin{array}{l}
2I_B \\
-2(I_B+1) \\
\end{array}
\right.
\qquad \hbox{for} \qquad I=
\left\{\begin{array}{l}
I_B+1/2 \\
I_B-1/2 \\
\end{array}
\right. \ \ ,
\label{fm1-7}
\end{equation}
and write the invariant $G$-matrix as
\begin{equation}
G(\bp, \bp^\prime; K, \omega)
=G^{I=I_B+1/2}_{BB}(\bp, \bp^\prime; K, \omega) \frac{1+P_\tau}{2}
+G^{I=I_B-1/2}_{BB}(\bp, \bp^\prime; K, \omega) \frac{1-P_\tau}{2}\ \ .
\label{fm1-8}
\end{equation}
We need to calculate 
\begin{equation}
X^{I \Omega}_B=\langle \chi_B \chi_\alpha |
\sum^5_{j=2} \omega^\Omega_{1,j}
\left(\frac{1 \pm P_\tau}{2}\right)_{1,j}
|\chi_B \chi_\alpha \rangle\ \ ,
\label{fm1-9}
\end{equation}
with the spin factors $\omega^{central}_{12}=1$,
$\omega^{ss}_{12}=(\bfsigma_1 \cdot \bfsigma_2)$,
$\omega^{LS}_{12}=\bfsigma_1+\bfsigma_2$,
and $\omega^{LS^{(-)}}_{12}=\bfsigma_1-\bfsigma_2$.
In \eq{fm1-9}, $\chi_\alpha$ is the spin-isospin wave
function of $\alpha$; i.e.,
$\phi_\alpha=\chi_\alpha \phi^{\rm space}_\alpha$.
Then we find that non-zero matrix elements are
\begin{equation}
X^{I\,central}_B=2\frac{2I+1}{2I_B+1}\ \ ,\qquad
X^{I\,LS}_B=X^{I\,LS^{(-)}}_B=2\frac{2I+1}{2I_B+1} \bfsigma_1 \ \ .
\label{fm1-10}
\end{equation}
On the other hand, the spatial part is calculated in Appendix
(B6) of Ref.\,\cite{2al}. It is convenient to express this
formula as
\begin{eqnarray}
& & V^{\rm space}(\bq_f, \bq_i)=\langle\,\delta(\bX_G) e^{i \bq_f \cdot \br}
\phi^{\rm space}_\alpha\,|\,G^{\rm space}\,|\,1 \cdot
e^{i \bq_i \cdot \br} \phi^{\rm space}_\alpha\,\rangle \nonumber \\
& & =e^{-\frac{3}{32\nu}k^2}
\left(\frac{2(1+\xi)^2}{3\pi \nu}\right)^{\frac{3}{2}}
\int d \bp\,\exp \left\{-\frac{2(1+\xi)^2}{3\nu}
\left(\bp-\frac{1+4\xi}{4(1+\xi)}\bq\right)^2\right\} \nonumber \\
& & \qquad \times G^{\rm space}\left(\bp+\frac{1}{2}\bk, \bp-\frac{1}{2}\bk;
(1+\xi)|\bq-\bp|, \omega \right)\ \ ,
\label{fm1-11}
\end{eqnarray}
where
\begin{equation}
\bk=\bq_f-\bq_i\ \ ,\qquad \bq=\frac{1}{2}\left(\bq_f+\bq_i\right)\ \ ,
\label{fm1-12}
\end{equation}
are the momentum transfer and the local momentum of the $B_8 \alpha$ system.
By assuming $g^I_0$, $h^I_0\,i\widehat{\bn}$ and
$h^I_-\,i\widehat{\bn}$ for $G^{\rm space}$,
we finally obtain 
\begin{eqnarray}
& & V(\bq_f, \bq_i) = \sum_{I \Omega} 2\left(\frac{2I+1}{2I_B+1}\right)
V^{I \Omega}(\bq_f, \bq_i)\ \ ,\nonumber \\
& & V^{I\,\left({C \atop LS}\right)}(\bq_f, \bq_i) \nonumber \\
& & = e^{-\frac{3}{32\nu}k^2}
\left(\frac{2(1+\xi)^2}{3\pi \nu}\right)^{\frac{3}{2}}
\int d \bp\,\exp \left\{-\frac{2(1+\xi)^2}{3\nu}
\left(\bp-\frac{1+4\xi}{4(1+\xi)}\bq\right)^2\right\} \nonumber \\
& & \qquad \times \left\{ \begin{array}{c}
g^I_0 \left(\bp+\frac{1}{2}\bk, \bp-\frac{1}{2}\bk;
(1+\xi)|\bq-\bp|, \omega \right) \\
h^I \left(\bp+\frac{1}{2}\bk, \bp-\frac{1}{2}\bk;
(1+\xi)|\bq-\bp|, \omega \right)
i \widehat{[\bp \times \bk]}\cdot \bfsigma_1 \\
\end{array} \right\}\ \ ,
\label{fm1-13}
\end{eqnarray} 
where $h^I=h^I_0+ h^I_-$ and $\widehat{[\bp \times \bk]}
=[\bp \times \bk]/|\,[\bp \times \bk]\,|$.

It should be noted that the c.m. momentum of the two interacting
particles, $\bK=(1+\xi)(\bq-\bp)$, in Eqs.\,(\ref{fm1-11}) and (\ref{fm1-13})
implies that the local momentum $\bq=(\bq_f+\bq_i)/2$ in \eq{fm1-12} plays
the role of the incident momentum of the first baryon $B$ from
\eq{fm1-3}, and $\bp$ the local momentum of the two-particle system,
which is now an integral variable. This is a consequence of fixing
the c.m. motion of the $B_8 \alpha$ system as in \eq{fm1-1}, and is
a special situation of the direct and knock-on (exchange) terms.
The $G$-matrix depends only on the magnitude $K=|\bK|$, since we
make an angular average in the $G$-matrix calculation \cite{GMAT,SPLS}.
The $G$-matrix value is, therefore, specified by $K$ and $\omega$,
or alternatively by the incident momentum $q_1=|\bq_1|$ and the relative
momentum $q=|\bq|$ between $B$ and $N$;
i.e., $G(\bp, \bp^\prime; K, \omega)=
G(\bq+\bk/2, \bq-\bk/2; q_1, q)$ with $\bk=\bp-\bp^\prime$ and
$\bq=(\bp+\bp^\prime)/2$. If $q_1$ and $q$ are specified, $K$ is
determined by the angular averaging, and $q_2$ is determined
from $q_1$ and $K$. Then the starting energy $\omega$ is determined
as the sum of the relative kinetic energy of two particles and 
the s.p. potentials for $B_8$ and $N$ at $q_1$ and $q_2$, respectively.
We therefore choose $q_1=|\bq_f+\bq_i|/2$ and
$q=|\bp|$ in the $G$-matrix in Eqs.\,(\ref{fm1-11}) and (\ref{fm1-13}).
In order to carry out this rather involved calculation,
we will develop in the next subsection some kind of transformation
formula for the rearrangement of relative momenta
in the partial-wave components of nonlocal kernels.

\vspace{-3mm}

\subsection{A transformation formula of the nonlocal kernel
in the momentum representation}

\vspace{-3mm}

The folding formula in \eq{fm1-11} implies that
expressing the $G$-matrix interaction $G^C(\bp, \bp^\prime)$
and $G^{LS}(\bp, \bp^\prime)\,i \widehat{[\bp^\prime \times \bp]}
\cdot \bS$ with the subsidiary momentum variables
\begin{eqnarray}
\bk & = & \bp-\bp^\prime\ \ ,\qquad
\qquad \bp=\bq+\frac{1}{2}\bk\ \ ,\nonumber \\
\bq & = & \frac{1}{2}(\bp+\bp^\prime)\ \ ,
\qquad \bp^\prime=\bq-\frac{1}{2}\bk\ \ ,
\label{fm2-1}
\end{eqnarray}
is convenient for the $\alpha$-cluster folding.
We express these kernels using the calligraphic letters; i.e.,
\begin{eqnarray}
& & \CG^C(\bk, \bq)=G^C(\bp, \bp^\prime)\ \ ,\nonumber \\
& & \CG^{LS}(\bk, \bq)\,i \widehat{[\bq \times \bk]}
\cdot \bS=G^{LS}(\bp, \bp^\prime)
\,i \widehat{[\bp^\prime \times \bp]}\cdot \bS\ \ .
\label{fm2-2}
\end{eqnarray}
By using this notation and the $q_1$, $q$ notation
for $K$, $\omega$, discussed in the last
subsection, $G^{\rm space}$ in \eq{fm1-11},
for example, becomes $\CG^{\rm space}(\bk, \bp; q_1, p)$ 
with $q_1=|\bq_f+\bq_i|/2$. In the following, we omit
the argument $K$, $\omega$ or $q_1$, $q$ for simplicity,
unless the dependence becomes crucial for the
relationship under consideration.
After making partial wave decomposition in \eq{fm2-2},
we can easily carry out the integral over $\bp$.
The resultant expression is denoted by $\CV^C(\bk, \bq)$
and by $\CV^{LS}(\bk, \bq)\,i\widehat{[\bq \times \bk]}\cdot \bS$.
The desired Born kernels, $V^C(\bq_f, \bq_i)$ and
$V^{LS}(\bq_f, \bq_i)\,i\widehat{[\bq_i \times \bq_f]}\cdot \bS$,
are obtained from another transformation related to \eq{fm1-12}.

Our task is, therefore, to relate the partial-wave components
$\CG^C_\lambda(k, q)$ and $G^C_\ell(p, p^\prime)$,
and also $\CG^{LS}_\lambda(k, q)$ and $G^{LS}_\ell(p, p^\prime)$,
which are defined through
\begin{eqnarray}
& & \CG^C(\bk, \bq)=\sum^\infty_{\lambda=0}
(2 \lambda+1) \CG^C_\lambda(k, q)
P_\lambda (\widehat{\bk}\cdot \widehat{\bq})\ \ ,
\nonumber \\
& & G^C(\bp, \bp^\prime)=\sum^\infty_{\ell=0}
(2 \ell+1) G^C_\ell(p, p^\prime)
P_\ell (\widehat{\bp}\cdot \widehat{\bp}^\prime)\ \ ,
\nonumber \\
& & \CG^{LS}(\bk, \bq)=\sum^\infty_{\lambda=1}
(2 \lambda+1) \CG^{LS}_\lambda(k, q)
P^1_\lambda (\widehat{\bk}\cdot \widehat{\bq})\ \ ,
\nonumber \\
& & G^{LS}(\bp, \bp^\prime)=\sum^\infty_{\ell=1}
(2 \ell+1) G^{LS}_\ell(p, p^\prime)
P^1_\ell (\widehat{\bp}\cdot \widehat{\bp}^\prime)\ \ .
\label{fm2-3}
\end{eqnarray}
We generalize the transformation \eq{fm2-1} as
\begin{eqnarray}
\left(\begin{array}{c}
\bp \\
\bp^\prime \\
\end{array} \right)
=\left(\begin{array}{cc}
\frac{1}{2} & 1 \\
-\frac{1}{2} & 1 \\
\end{array} \right)
\left(\begin{array}{c}
\bk \\
\bq \\
\end{array} \right)
=\left(\begin{array}{cc}
\alpha & \beta \\
\gamma & \delta \\
\end{array} \right)
\left(\begin{array}{c}
\bk \\
\bq \\
\end{array} \right)\ \ ,
\label{fm2-4}
\end{eqnarray}
with $\alpha \delta - \beta \gamma=1$.
A simple calculation gives
\begin{equation}
\CG^{C}_\lambda (k, q)=\sum^\infty_{\ell=0} (2\ell+1)
\frac{1}{2} \int^1_{-1} dx~\frac{G^C_\ell(p, p^\prime)}
{(p p^\prime)^\ell} P_\lambda(x)\,g_\ell(k, q; x)\ \ ,
\label{fm2-4-2}
\end{equation}
where $g_\ell(k, q; x)=(p p^\prime)^\ell
P_\ell(\widehat{\bp}\cdot \widehat{\bp}^\prime)$ with
\begin{equation}
p=\sqrt{\alpha^2 k^2+\beta^2 q^2+2\alpha \beta kqx}
\ \ ,\qquad p^\prime=\sqrt{\gamma^2 k^2
+\delta^2 q^2+2\gamma \delta kqx}\ \ .
\label{fm2-5}
\end{equation}
The transformation of the $LS$ components is similarly
carried out by using the orthogonality relations of
$P^1_\ell(x)$ and
\begin{eqnarray}
& & \left(\frac{1}{\sin \theta}\right) P^1_\ell(\cos \theta)
=\sum_{\ell^\prime=\ell-1, \ell-3,\cdots, 1~{\rm or}~0}
(2\ell^\prime+1)\,P_{\ell^\prime}(\cos \theta)\ \ ,\nonumber \\
& & P_{\lambda-1}(x)-P_{\lambda+1}(x)=\frac{2\lambda+1}
{\lambda (\lambda+1)} \sqrt{1-x^2}\,P^1_\lambda(x)\ \ .
\label{fm2-6}
\end{eqnarray}
For the numerical integration over $x$ in \eq{fm2-4-2} etc.,
it is convenient to use the spline interpolation
\begin{equation}
G^{\Omega}_\ell (p, p^\prime)=\sum_{i, j}
S_i(p)\,S_j(p^\prime)\,G^{\Omega}_\ell(p_i, p_j)\ \ ,
\label{fm2-7}
\end{equation}
since the G-matrix calculation itself is very much 
time consuming.
After all, we obtain the following formula for the
transformation of the partial-wave components:
\begin{eqnarray}
& & \CG^\Omega_\lambda(k, q)=\sum_{i,j}\sum^\infty_{\ell=0
~{\rm or}~1}(2\ell+1)\,F^{\Omega\,\lambda \ell}_{i,j}
\,G^\Omega_\ell(p_i, p_j)\qquad
(\Omega=C,~LS)\ \ ,\nonumber \\
& & F^{C\,\lambda \ell}_{i,j}
=\frac{1}{2}\int^1_{-1} dx~\frac{S_i(p) S_j(p^\prime)}
{(pp^\prime)^\ell} P_\lambda(x)\,g_\ell(k,q;x)\ \ ,\nonumber \\
& & F^{LS\,\lambda \ell}_{i,j}
=\frac{kq}{2\lambda+1}
\sum_{\ell^\prime=\ell-1, \ell-3,\cdots, 1~{\rm or}~0}
(2\ell^\prime+1)\,\frac{1}{2}\int^1_{-1} dx
~\frac{S_i(p) S_j(p^\prime)}
{(pp^\prime)^{\ell^\prime+1}} \nonumber \\
& & \qquad \times \left[ P_{\lambda-1}(x)-P_{\lambda+1}(x) \right]
\,g_{\ell^\prime}(k,q;x)\ \ ,
\label{fm2-8}
\end{eqnarray}
where
\begin{eqnarray}
& & g_\ell(k, q; x)=(p p^\prime)^\ell
P_\ell(\widehat{\bp}\cdot \widehat{\bp}^\prime)
\nonumber \\
& & =\sum_{\ell_1+\ell_2=\ell}\sum_{{\ell_1}^\prime
+{\ell_2}^\prime=\ell}
(-1)^{\ell_2+{\ell_1}^\prime}
\frac{(2\ell+1)\,!}{\sqrt{(2\ell_1)\,!\,(2\ell_2)\,!
\,(2{\ell_1}^\prime)\,!\,(2{\ell_2}^\prime)\,!}}
\alpha^{\ell_1} \beta^{\ell_2} \gamma^{{\ell_1}^\prime}
\delta^{{\ell_2}^\prime}
\nonumber \\
& & \times k^{\ell_1+{\ell_1}^\prime}
q^{\ell_2+{\ell_2}^\prime}
\sum_{\ell^\prime} \langle \ell_1 0 {\ell_1}^\prime 0
|\ell^\prime 0 \rangle \langle \ell_2 0 {\ell_2}^\prime 0|
\ell^\prime 0 \rangle \left\{\begin{array}{ccc}
\ell_1 & \ell_2 & \ell \\
{\ell_2}^\prime & {\ell_1}^\prime & \ell^\prime \\
\end{array} \right\} P_{\ell^\prime}(x)\ \ .
\label{fm2-9}
\end{eqnarray}
The summation over $\ell$ in \eq{fm2-8} is from $\ell=0$ for
$\Omega=C$ and from $\ell=1$ for $\Omega=LS$.

It is important to note some symmetries possessed
by $\CG^\Omega_\lambda(\bk, \bq)$.
From the time-reversal symmetry,
the $G$-matrix $G^\Omega (\bp, \bp^\prime)$ is symmetric for
the interchange of $\bp$ and $\bp^\prime$. Since this interchange
corresponds to $\bk \rightarrow -\bk$, the transformed
partial-wave components, $\CG^C_\lambda(k, q)$
and $\CG^{LS}_\lambda(k, q)$, are no-zero
only for $\lambda=0~,2~,4~,\cdots$ and $\lambda=1~,3~,5~,\cdots$,
respectively. For the coefficients in \eq{fm2-4}, we can
show $g_\ell(2q, k/2; x)=(-1)^\ell g_\ell(k, q; x)$ in \eq{fm2-9}.
This property implies that, if $G^\Omega_\ell(p, p^\prime)$ is
transformed to $\CG^\Omega_\lambda(k, q)$, then
$(-1)^\ell G^\Omega_\ell(p, p^\prime)$ is transformed
to $\CG^\Omega_\lambda(2q, k/2)$. In the later application
of the present formalism to $n \alpha$ resonating-group method (RGM),
we will find that the knock-on term is obtained by simply replacing 
$\CG^\Omega_\lambda(k, q)$ to $\CG^\Omega_\lambda(2q, k/2)$ in
the direct term. In fact, the knock-on term is
already included even in the hyperon $\alpha$ system
as in \eq{fm1-4}, which is the strangeness exchange term
of the hyperon-nucleon interaction. In other words, the
direct and knock-on terms are treated on the equal
footing in the present formalism to deal with the invariant
$G$-matrix interaction. For the transformation from
$\CV^\Omega(\bk, \bq)$ to $V^\Omega(\bq_f, \bq_i)$,
we find that the latter Born kernel is symmetric with
respect to the interchange of $q_f$ and $q_i$, which is the
consequence of Eqs.\,(\ref{fm2-5}) and (\ref{fm2-8}) for
the transformation coefficients (see \eq{fm1-12})
\begin{eqnarray}
\left(\begin{array}{c}
\bk \\
\bq \\
\end{array} \right)
=\left(\begin{array}{cc}
1 & -1 \\
\frac{1}{2} & \frac{1}{2} \\
\end{array} \right)
\left(\begin{array}{c}
\bq_f \\
\bq_i \\
\end{array} \right)\ \ .
\label{fm2-10}
\end{eqnarray}

\vspace{-3mm}

\subsection{Partial-wave expansion of the $B_8 \alpha$ Born kernel}

\vspace{-3mm}

If we use partial-wave decomposition in \eq{fm2-3} and the similar expansions
for $\CV^\Omega(\bk, \bq)=V^\Omega(\bq_f, \bq_i)$, the folding
formula \eq{fm1-11} becomes very simple. For both of the
$\Omega=C$ and $LS$ terms, it is given by
\begin{eqnarray}
& & \CV^\Omega_\lambda (k, q)=\exp \left\{-\frac{3}{32\nu}k^2
-\frac{2}{3\nu}\left(\frac{1}{4}+\xi\right)^2 q^2\right\}
\left(\frac{2(1+\xi)^2}{3\pi \nu}\right)^{\frac{3}{2}} 4\pi
\nonumber \\
& & \times \int^\infty_0 p^2 dp~\exp \left\{-\frac{2(1+\xi)^2}{3\nu}p^2
\right\}\,i_\lambda\left(\frac{(1+\xi)(1+4\xi)}{3\nu}pq\right)
~\CG^\Omega_\lambda(k, p)\ \ ,\nonumber \\
\label{fm3-1}
\end{eqnarray}
where $i_\lambda(x)=i^\lambda j_\lambda(-ix)$ is the spherical
Bessel function of the imaginary argument.
For the proof of the $LS$-term folding, we use a simple formula
\begin{equation}
P^1_\ell (\widehat{\bk}\cdot \widehat{\bq})
~i \widehat{\left[\bq \times \bk\right]}_\nu
=(-1)^\ell \frac{4\pi}{\sqrt{3}} \sqrt{\frac{\ell (\ell+1)}
{~2\ell+1~}}~\left[Y_\ell(\widehat{\bk}) Y_\ell(\widehat{\bq})
\right]_{1\nu}\ \ ,
\label{fm3-2}
\end{equation}
which is derived by using
\begin{equation}
P^1_{\ell+1}(x)-P^1_{\ell-1}(x)=(2\ell+1)\,\sqrt{1-x^2}\,P_\ell(x)\ \ ,
\label{fm3-3}
\end{equation}
and
\begin{eqnarray}
& & P_\ell (\widehat{\bk}\cdot \widehat{\bq})
~i \left[\bq \times \bk\right]_\nu
=(-1)^\ell \frac{4\pi}{\sqrt{3}}\frac{kq}{(2\ell+1)}
\left\{\sqrt{\frac{(\ell+1)(\ell+2)}{2\ell+3}}
\right.\nonumber \\
& & \left. \times 
\left[Y_{\ell+1}(\widehat{\bq}) Y_{\ell+1}(\widehat{\bk})\right]_{1\nu}
-\sqrt{\frac{(\ell-1)\ell}{2\ell-1}}
\left[Y_{\ell-1}(\widehat{\bq}) Y_{\ell-1}(\widehat{\bk})\right]_{1\nu}
\right\}\ \ .
\label{fm3-4}
\end{eqnarray}

The final step to derive the partial-wave components
$V^\Omega_\ell(q_f, q_i)$ in
\begin{eqnarray}
V(\bq_f, \bq_i) & = & V^C(\bq_f, \bq_i)+V^{LS}(\bq_f, \bq_i)
~i \widehat{[\bq_i \times \bq_f]}\cdot \bS\ \ ,\nonumber \\
V^C(\bq_f, \bq_i) & = & \sum^\infty_{\ell=0}
(2\ell+1)\,V^C_\ell(q_f, q_i)
\,P_\ell (\widehat{\bq}_f\cdot \widehat{\bq}_i)\ \ ,\nonumber \\
V^{LS}(\bq_f, \bq_i) & = & \sum^\infty_{\ell=1}
(2\ell+1)\,V^{LS}_\ell(q_f, q_i)
\,P^1_\ell (\widehat{\bq}_f\cdot \widehat{\bq}_i)\ \ ,
\label{fm3-5}
\end{eqnarray}
is carried out by using the transformation formula
in the preceding subsection with the coefficients
in \eq{fm2-10}.
In the $jj$-coupling scheme, we also need $V^J_\ell(q_f, q_i)$
in the partial-wave expansion
\begin{eqnarray}
V(\bq_f, \bq_i)=4\pi \sum_{J \ell} V^J_\ell(q_f, q_i)
\sum_M \CY_{(\ell\frac{1}{2})JM}\left(\widehat{\bq}_f; {\rm spin}\right)
\,\CY^*_{(\ell\frac{1}{2})JM}\left(\widehat{\bq}_i; {\rm spin}\right)
\ \ ,\nonumber \\
\label{fm3-6}
\end{eqnarray}
which is given by
\begin{eqnarray}
& & \ \hspace{-5mm} V^J_\ell(q_f, q_i)=V^C_\ell(q_f, q_i)
+V^{LS}_\ell(q_f, q_i)~\langle \bL \cdot \bS \rangle_{J\ell}
\ \ ,\nonumber \\
& & \ \hspace{-5mm} \langle \bL \cdot \bS \rangle_{J\ell}
=\frac{1}{2}\left[J(J+1)-\ell(\ell+1)-\frac{3}{4}\right]
=\left\{\begin{array}{c}
\frac{\ell}{2} \\
-\frac{\ell+1}{2} \\
\end{array} \right.
\quad \hbox{for} \quad
J=\left\{\begin{array}{c}
\ell+\frac{1}{2} \\
\ell-\frac{1}{2} \\
\end{array} \right.
\ \ .\nonumber \\
\label{fm3-7}
\end{eqnarray}
In \eq{fm3-6}, $\CY_{(\ell\frac{1}{2})JM}\left(\widehat{\bq};
{\rm spin}\right)=[Y_\ell(\widehat{\bq})\,\chi_{\frac{1}{2}}]_{JM}$ is
the angular-spin wave function of the
$B_8 \alpha$ system. For the proof, we again use \eq{fm3-2}.

In summary, the Born kernel of the $B_8 \alpha$ system
is obtained by using Eqs. (\ref{fm2-8}) and (\ref{fm3-1}),
starting from
\begin{eqnarray}
& & \ \hspace{-10mm} G^C_\ell(p, p^\prime)=\frac{1}{2} \sum_{IJS}
\left(\frac{2I+1}{2I_B+1}\right)\left(\frac{2J+1}{2\ell+1}\right)
G^{IJ}_{S\ell, S\ell}(p, p^\prime)\quad (\ell=0,~1,~2,\cdots)
\ \ ,\nonumber \\
& & \ \hspace{-10mm} G^{LS}_\ell(p, p^\prime)=\sum_I
\left(\frac{2I+1}{2I_B+1}\right)
\left\{-\frac{1}{\ell(\ell+1)} G^{I\,\ell}_{1\ell, 1\ell}(p, p^\prime)
-\frac{2\ell-1}{\ell(2\ell+1)} G^{I\,\ell-1}_{1\ell, 1\ell}(p, p^\prime)
\right.\nonumber \\
& & \left.\ \hspace{-10mm}
+\frac{2\ell+3}{(\ell+1)(2\ell+1)} G^{I\,\ell+1}_{1\ell, 1\ell}(p, p^\prime)
+\frac{1}{\sqrt{\ell(\ell+1)}}\left[G^{I\,\ell}_{1\ell, 0\ell}(p, p^\prime)
+G^{I\,\ell}_{0\ell, 1\ell}(p, p^\prime)\right] \right\}
\nonumber \\
& & \hspace{80mm} (\ell=1,~2,~3,\cdots)\ \ .
\label{fm3-8}  
\end{eqnarray}
For $B_8=N$, we should multiply the factor 2 to include the knock-on term.
We first use \eq{fm2-8} with the coefficients \eq{fm2-4}
and transform the above $G^\Omega_\ell(p, p^\prime)$
to $\CG^\Omega_\lambda(k, q)$. Then, the $\alpha$-cluster folding
by \eq{fm3-1} yields $\CV^\Omega_\lambda(k, q)$ from
$\CG^\Omega_\lambda(k, q)$. The second transformation
from $\CV^\Omega_\lambda(k, q)$ to $V^\Omega_\lambda(q_f, q_i)$ is
carried out with the coefficient \eq{fm2-10}. 
The selection of $q_1$ and $q$ in $G^\Omega(p, p^\prime; q_1, q)$
is now almost apparent. We choose $q_1$ as $q$ in the folding
formula \eq{fm3-1}. The relative momentum $q$ is actually $p$ in
\eq{fm3-1}. Therefore, the nest structure $\CV^\Omega_\lambda(k, q)
\supset \CG^\Omega_\lambda(k,p;q_1=q,q=p)$ and
$\CG^\Omega_\lambda(k,q;q_1,q) \supset G^\Omega_\ell(p, p^\prime; q_1, q)$
is incorporated in the computer code.
Namely, we first specify $k$ and $q$ in \eq{fm3-1}.
Then $\CG^\Omega_\lambda(k,p;q_1=q,q=p)$ is generated
from $\CG^\Omega_\lambda(k,q;q_1,q)$ for a general $q$,
which is obtained from the transformation formula, \eq{fm2-8},
by using the complete off-shell $G$-matrix
$G^\Omega_\ell(p_i, p_j; q_1, q)$.

\vspace{-3mm}

\subsection{Wigner transform}

\vspace{-3mm}

The present formalism is convenient to calculate the Wigner transform
$V^\Omega_W(\br, \bq)$, since they are essentially
the Fourier transform of $\CV^\Omega (\bk, \bq)$. We define these through
\begin{eqnarray}
V_W(\br, \bq) & = & \frac{1}{(2\pi)^3} \int d \bk~e^{i\bk \cdot \br}
\,V(\bq+\bk/2, \bq-\bk/2)
\nonumber \\
& = & \frac{1}{(2\pi)^3} \int d \bk~e^{i\bk \cdot \br}
\,\left\{\CV^C(\bk, \bq)+\CV^{LS}(\bk, \bq)
\,i\widehat{[\bq \times \bk]}\cdot \bS\right\} \nonumber \\
& = & V^C_W(\br, \bq)+V^{LS}_W(\br, \bq)\,[\br \times \bq]\cdot \bS\ \ .
\label{fm4-1}
\end{eqnarray}
Note that the $LS$ term is defined by $\bL_W\cdot \bS \equiv [\br \times \bq]
\cdot \bS$, instead of $i \widehat{[\br \times \bq]}\cdot \bS$.
If we apply the partial-wave expansion
\begin{eqnarray}
V^C_W(\br, \bq) & = & \sum^\infty_{\lambda=0}
(2\lambda+1)\,V^C_{W \lambda}(r, q)
\,P_\lambda (\cos \varphi)\ \ ,\nonumber \\
V^{LS}_W(\br, \bq) & = & \sum^\infty_{\lambda=1}
(2\lambda+1)\,V^{LS}_{W \lambda}(r, q)
\,\left(\frac{1}{\sin \varphi}\right)
P^1_\lambda (\cos \varphi)\ \ ,
\label{fm4-2}
\end{eqnarray}
with $\cos \varphi=(\widehat{\br}\cdot \widehat{\bq})$,
we can easily derive the partial-wave components as
\begin{eqnarray}
V^C_{W \lambda}(r, q) & = & \frac{4\pi}{(2\pi)^3}
\int^\infty_0 k^2 dk~i^\lambda\,j_\lambda(kr)\,\CV^C_\lambda(k,q)
\ \ ,\nonumber \\
V^{LS}_{W \lambda}(r, q) & = & \frac{1}{qr}
\,\frac{4\pi}{(2\pi)^3}
\int^\infty_0 k^2 dk~i^{\lambda-1}\,j_\lambda(kr)\,\CV^{LS}_\lambda(k,q)
\ \ .
\label{fm4-3}
\end{eqnarray}
Here we use \eq{fm3-2} again for the derivation of the $LS$ term.
Since $V^C_{W \lambda}(r, q) \ne 0$ for $\lambda=0,~2,~4,\cdots$
and $V^{LS}_{W \lambda}(r, q) \ne 0$ for $\lambda=1,~3,~5,\cdots$
(see Subsec. 2.2), the $\lambda=0$ and $\lambda=1$ terms become
the leading terms for the central and $LS$ Wigner transform,
respectively. In fact, in the $q=0$ case, we find that only these
leading terms survive for the zero-momentum Wigner transform.
It is, therefore, a good approximation to retain only $\lambda=0$
and $\lambda=1$ terms in \eq{fm4-2}:
\begin{eqnarray}
V^C_W(\br, \bq) & \sim & \frac{4\pi}{(2\pi)^3}
\int^\infty_0 k^2 dk~j_0(kr)\,\CV^C_0(k, q)
\ \ ,\nonumber \\
V^{LS}_W(\br, \bq) & \sim & \frac{4\pi}{(2\pi)^3}
\int^\infty_0 k^2 dk~\frac{1}{r}\,j_1(kr)\,\frac{3}{q}\,\CV^{LS}_1(k, q)
\ \ ,
\label{fm4-4}
\end{eqnarray}
which we use throughout this paper.

The zero-momentum Wigner transform is a convenient tool to estimate
the strength of the interaction. From the folding formula \eq{fm3-1},
we can calculate $\CV^C_0(k, 0)$ and $(3/q)\CV^{LS}_1(k, q)|_{q=0}$.
This process yields
\begin{eqnarray}
V^C_W(\br, 0) & = & \frac{2}{\pi}
\int^\infty_0 k^2 dk~e^{-\frac{3}{32\nu}k^2}\,j_0(kr)\nonumber \\
& & \times \left(\frac{2(1+\xi)^2}{3\pi \nu}\right)^{\frac{3}{2}}
\int^\infty_0 q^2 dq~e^{-\frac{2(1+\xi)^2}{3\nu}q^2}
\,\CG^C_0(k, q)\ \ ,\nonumber \\
V^{LS}_W(\br, 0) & = & \frac{(1+\xi)(1+4\xi)}{3\nu}\,\frac{2}{\pi}
\int^\infty_0 k^2 dk~e^{-\frac{3}{32\nu}k^2}\,\frac{1}{r}
\,j_1(kr)\nonumber \\
& & \times \left(\frac{2(1+\xi)^2}{3\pi \nu}\right)^{\frac{3}{2}}
\int^\infty_0 q^3 dq~e^{-\frac{2(1+\xi)^2}{3\nu}q^2}
\,\CG^{LS}_1(k, q)\ \ .\nonumber \\
\label{fm4-5}
\end{eqnarray}

For the $LS$ component, it is easy to introduce another
approximation, which gives a simple factor for the $LS$
strength, similar to the Scheerbaum's factor \cite{SC76}.
For this purpose, we express $\CG^{LS}_1(k, q)$ in the original
form
\begin{eqnarray}
\CG^{LS}_1(k, q) & = & \frac{kq}{3}
\sum^\infty_{\ell=1} (2\ell+1) \frac{1}{2}
\int^1_{-1} dx~[1-P_2(x)]\,\frac{G^{LS}_\ell(p, p^\prime)}{p p^\prime}
\nonumber \\
& & \times \left(\frac{1}{\sin \theta}\right) P^1_\ell(\cos \theta)
\ \ ,
\label{fm4-6}
\end{eqnarray}
where $\cos \theta=(\widehat{\bp}\cdot \widehat{\bp}^\prime)$
with $\bp=\bq+\bk/2$, $\bp^\prime=\bq-\bk/2$
and $x=(\widehat{\bk}\cdot \widehat{\bq})$.
We neglect the $k$ dependence except for the front factor $kq/3$ and
set $k=0$. Then, $p,~p^\prime \rightarrow q$, $\cos \theta \rightarrow 1$,
and the $x$ integral can be performed. By further using 
$P^\prime_\ell(1)=\ell(\ell+1)/2$, we obtain
\begin{eqnarray}
\CG^{LS}_1(k, q) & \sim & \frac{kq}{3}
~\sum^\infty_{\ell=1} (2\ell+1)\,\frac{1}{q^2}
\,G^{LS}_\ell(q, q)\,\frac{\ell(\ell+1)}{2}\nonumber \\
& = & \left(\frac{2k}{3q}\right)\,\frac{1}{4}~\sum^\infty_{\ell=1}
\ell(\ell+1)(2\ell+1)\,G^{LS}_\ell(q, q)\ \ .
\label{fm4-7}
\end{eqnarray}
If we set
\begin{eqnarray}
\ \hspace{-10mm} G(q) & = & \frac{1}{4}~\sum^\infty_{\ell=1}
\ell(\ell+1)(2\ell+1)\,G^{LS}_\ell(q, q)\nonumber \\
\ \hspace{-10mm} & = & \frac{1}{4}~\sum_{IJ} \left(\frac{2I+1}{2I_B+1}\right)
(2J+1) \left\{-G^{IJ}_{1\,J,1\,J}(q,q)-(J+2) G^{IJ}_{1\,J+1,1\,J+1}(q,q)
\right. \nonumber \\
\ \hspace{-10mm} & & \left. +(J-1) G^{IJ}_{1\,J-1,1\,J-1}(q,q)
+\sqrt{J(J+1)} \left[G^{IJ}_{1\,J,0\,J}(q,q)+G^{IJ}_{0\,J,1\,J}(q,q)\right]
\right\}
\nonumber \\
\ \hspace{-10mm} & = & 2 \sum_{IJ} \left(\frac{2I+1}{2I_B+1}\right)
\left[ h^I_0(\bp, \bp^\prime)+h^I_-(\bp, \bp^\prime)\right]
\frac{1}{\sin \theta} \Bigr|_{\bp=\bp^\prime=\bq}\ \ ,
\label{fm4-8}
\end{eqnarray}
we can write $\CG^{LS}_1(k, q)\sim (2k/3q)G(q)$. Using this
and the integration formula
\begin{equation}
\frac{2}{\pi}\int^\infty_0 k^3 dk~e^{-\frac{3}{32\nu}k^2}
\,\frac{1}{r}\,j_1(kr)=\frac{8}{\sqrt{\pi}}
\,\left(\frac{8\nu}{3}\right)^{\frac{5}{2}}
\,e^{-\frac{8}{3}\nu r^2}\ \ ,
\label{fm4-9}
\end{equation}
we find 
\begin{eqnarray}
V^{LS}_W(\br, 0) & \sim & \frac{2}{3}
\,\frac{(1+\xi)(1+4\xi)}{3\nu}\,
\frac{8}{\sqrt{\pi}}\,\left(\frac{8\nu}{3}\right)^{\frac{5}{2}}
\,e^{-\frac{8}{3}\nu r^2}\nonumber \\
& & \times \left(\frac{2(1+\xi)^2}{3\pi \nu}\right)^{\frac{3}{2}}
\int^\infty_0 q^2 dq~e^{-\frac{2(1+\xi)^2}{3\nu}q^2}
\,G(q)\ \ .
\label{fm4-10}
\end{eqnarray}
We can write \eq{fm4-10} in the way similar to the Scheerbaum's formula:
\begin{equation}
U(r)=-\frac{\pi}{2}\,S_B\,\frac{1}{r}\,\frac{d \rho(r)}{d r}
\,\bfell \cdot \bfsigma\ \ .
\label{fm4-11}
\end{equation}
For the $B_8 \alpha$ system, the density $\rho(r)$ of the $\alpha$-cluster
is calculated as
\begin{equation}
\rho(r)=\langle \phi_\alpha | \sum^4_{i=1} \delta (\br-\bfxi_i) |
\phi_\alpha \rangle
=4\left(\frac{8\nu}{3\pi}\right)^{\frac{3}{2}}\,e^{-\frac{8}{3}\nu r^2}
\ \ ,
\label{fm4-12}
\end{equation}
where $\bfxi_i=\bx_i-\bX_\alpha$ is the coordinate of the nucleon $i$, measured
from the c.m. of the $\alpha$-cluster.
Then, we find that the integral \eq{fm4-9} is nothing
but $(-\pi)(1/r)(d\rho(r)/dr)$. Thus \eq{fm4-11} becomes
\begin{equation}
U(r)=S_B\,\frac{8}{\sqrt{\pi}}\,\left(\frac{8\nu}{3}\right)^{\frac{5}{2}}
\,e^{-\frac{8}{3}\nu r^2}\,\bfell \cdot \bS\ \ .
\label{fm4-13}
\end{equation}
Similarly, we can write \eq{fm4-10} as
\begin{equation}
V^{LS}_W(\br, 0)=\widetilde{S}_B
\,\frac{8}{\sqrt{\pi}}\,\left(\frac{8\nu}{3}\right)^{\frac{5}{2}}
\,e^{-\frac{8}{3}\nu r^2}\ \ ,
\label{fm4-13-2}
\end{equation}
and obtain
\begin{equation}
\widetilde{S}_B=\frac{1+4\xi}{4\xi}\,\frac{1}{2\pi}
\,\frac{\xi}{1+\xi}\,\frac{8\pi^2}{3}\,
\left(\frac{2(1+\xi)^2}{3\pi \nu}\right)^{\frac{5}{2}}
\int^\infty_0 q^2 dq~e^{-\frac{2(1+\xi)^2}{3\nu}q^2}
\,G(q)\ \ .
\label{fm4-14}
\end{equation}
This expression corresponds to Eq.\,(50) of Ref.\,\cite{SPLS},
which is the Scheerbaum factor of finite nuclei derived from
$G$-matrix calculations:
\begin{equation}
S_B(q_1)=\frac{1}{2\pi}
\,\frac{\xi}{1+\xi}\,\frac{3}{(k_F)^3}\,(1+\xi)^3
\int^{q_{\rm max}}_0 dq~W(q_1, q)\,G(q)\ \ .
\label{fm4-15}
\end{equation}
The difference between Eqs.\,(\ref{fm4-14}) and (\ref{fm4-15}) is
the weight function,
\begin{equation}
W(q)=q^2\,e^{-\frac{2(1+\xi)^2}{3\nu}q^2} \longleftrightarrow
W(q_1, q)\ \ ,
\label{fm4-16}
\end{equation}
and an extra front factor $\frac{1+4\xi}{4\xi}$ in \eq{fm4-14}.
This enhancement factor appears, since in the standard $G$-matrix
calculation of single-particle potentials the c.m. motion
of the total system is not correctly treated.
In the $B_8 \alpha$ system, this approximation
affects the result appreciably, which is discussed in Ref.\,\cite{2aljj}.
The two weight functions in \eq{fm4-16} may seem to be fairly different,
since $W(q_1, q)$ with $q_1=0$ is given by
$W(0,q)=\theta (q-k_F/(1+\xi))$. However, this is not the case,
since in the small $q$ region (the low-energy region) $G(q)$ is
anyway very small. We can further approximate  
Eqs.\,(\ref{fm4-14}) and (\ref{fm4-15}), by calculating
$\bar{q}=\sqrt{\langle q^2 \rangle}$ with each weight function,
and by replacing $q$ in $G(q)$ with this $\bar{q}$.
The average momentum $\bar{q}$ is given by 
\begin{equation}
\bar{q}=\frac{3\sqrt{\nu}}{2}\,\left(\frac{1}{1+\xi}\right)
\quad \hbox{for} \quad W(q)\ ,\quad
\bar{q}=\frac{k_F}{\sqrt{3}}\,\left(\frac{1}{1+\xi}\right)
\quad \hbox{for} \quad W(0,q)\ \ .
\label{fm4-17}
\end{equation}
For $W(q)$ with $\nu=0.257~\hbox{fm}^{-1}$,
$\bar{q} \sim 0.38~\hbox{fm}^{-1}$, and for $W(0, q_1)$ with
$k_F=1.2~\hbox{fm}^{-1}$, $\bar{q} \sim 0.34~\hbox{fm}^{-1}$.
These values are very close to $q=\bar{k}/2=0.35~\hbox{fm}^{-1}$,
calculated from the Scheerbaum's estimation for
an average momentum transfer $\bar{k}\sim 0.7~\hbox{fm}^{-1}$. 
From this prescription, $\tilde{S}_B$ and $S_B(0)$ are given by
\begin{equation}
\tilde{S}_B=\frac{1+4\xi}{4\xi}\,\frac{1}{2\pi}
\,\frac{\xi}{1+\xi}\,\frac{1}{\bar{q}^2}\,G(\bar{q})\ \ ,
\qquad S_B(0)=\frac{1}{2\pi}
\,\frac{\xi}{1+\xi}\,\frac{1}{\bar{q}^2}\,G(\bar{q})\ \ .
\label{fm4-18}
\end{equation}

\begin{figure}[b]
\begin{center}
\begin{minipage}[h]{0.49\textwidth}
\includegraphics[width=\textwidth]{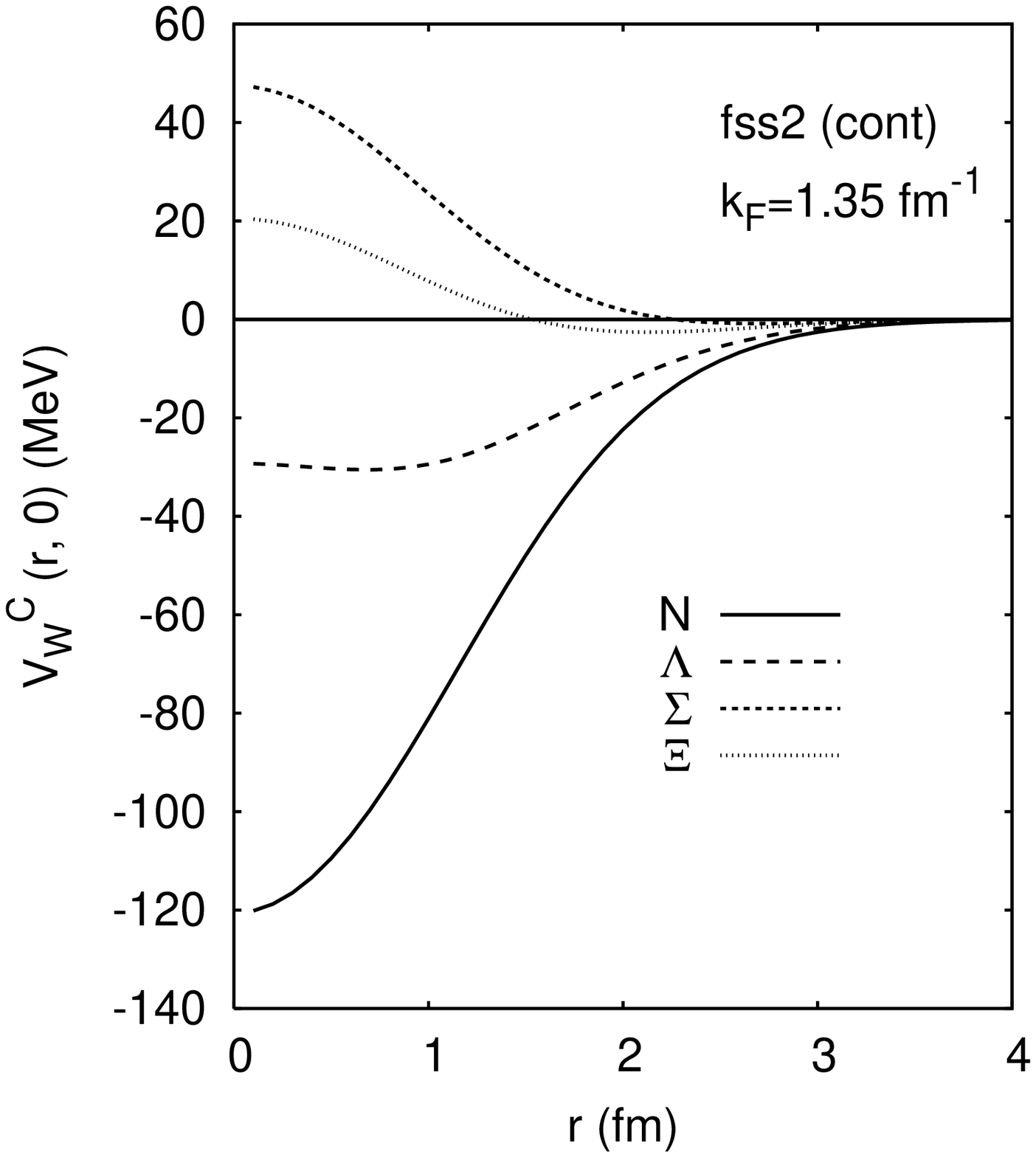}
\caption{The central component of the zero-momentum Wigner
transform ${V_W}^C(r ,0)$ for the $B_8 \alpha$ Born kernel,
calculated from the quark-model $G$-matrix
$B_8 B_8$ interactions by fss2.
The Fermi momentum used in the $G$-matrix calculation
is $k_F=1.35~\hbox{fm}^{-1}$ and the continuous choice is used
for intermediate spectra. 
The $(0s)^4$ shell-model wave function with the h.o. size
parameter $\nu=0.257~\hbox{fm}^{-2}$ is used for the $\alpha$ cluster.
}
\label{fig1a}
\end{minipage}
\hfill
\begin{minipage}[h]{0.49\textwidth}
\vspace{-42mm}
\includegraphics[width=\textwidth]{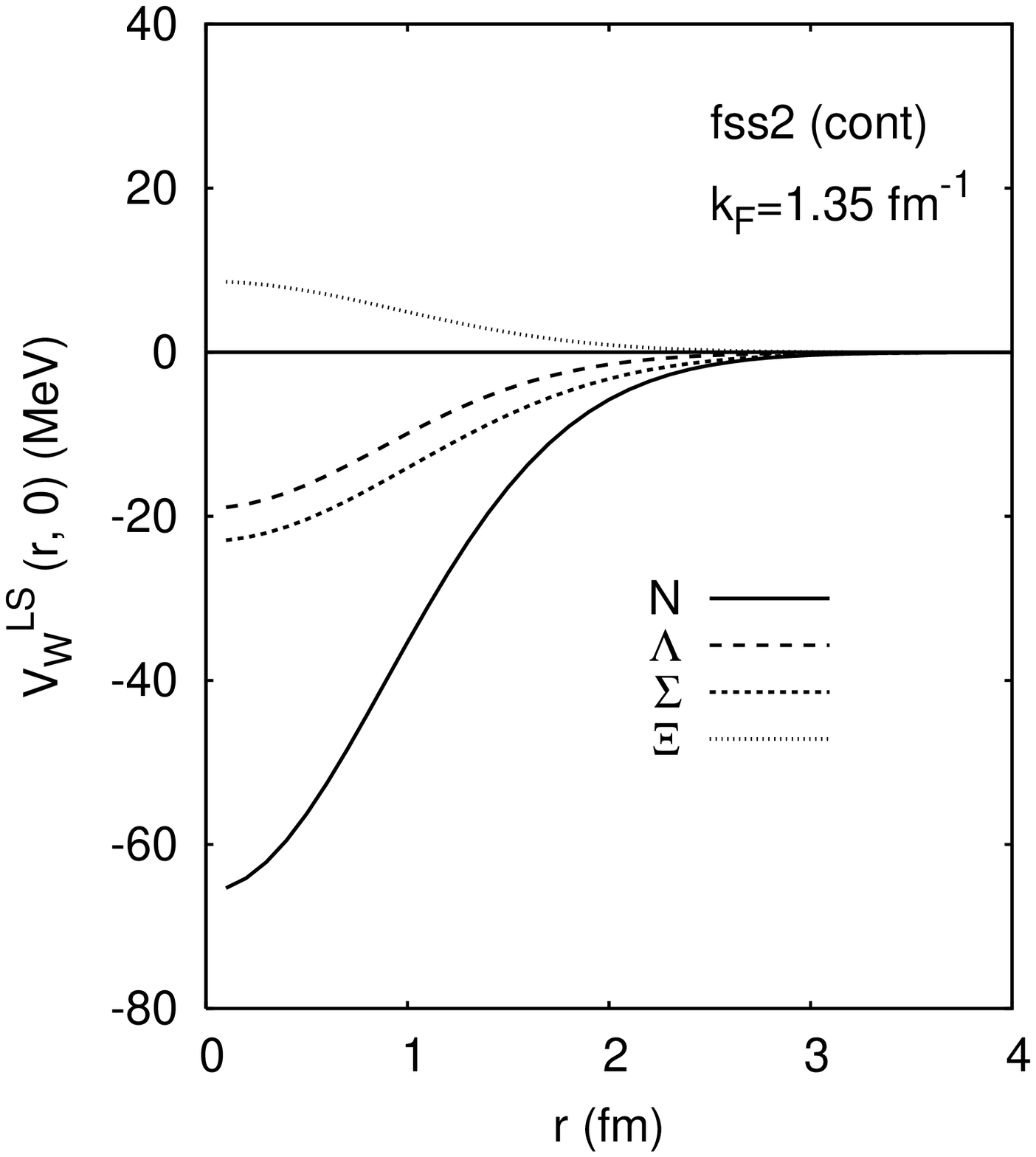}
\caption{
The same as Fig.~\protect\ref{fig1a}, but for
the $LS$ component ${V_W}^{LS}(r, 0)$.
}
\label{fig1b}
\end{minipage}
\end{center}
\end{figure}

\vspace{-3mm}

\section{Results and discussion}

\vspace{-3mm}

\subsection{$\Lambda \alpha$ interaction}

\vspace{-3mm}

In order to gain the overview of the zero-momentum Wigner transform
given in \eq{fm4-5}, we will show in Fig.\,\ref{fig1a} the
central ($V^C_W(\br, 0)$) and $LS$ ($V^{LS}_W(\br, 0)$) components
for the $B_8 \alpha$ interaction with $B_8=N,~\Lambda,~\Sigma$ and $\Xi$,
when the quark-model $G$-matrix $B_8 B_8$ interaction by fss2 are
employed with the continuous choice for intermediate spectra.
In this case, we choose $q_1=0$ and the relative momentum $q$ is
assigned to the integral mesh point $q$ in \eq{fm4-5}.
The Fermi-momentum used for the $G$-matrix calculation is
$k_F=1.35~\hbox{fm}^{-1}$, and the $(0s)^4$ shell-model wave
function with the h.o. size parameter $\nu=0.257~\hbox{fm}^{-2}$ is
used for the $\alpha$-cluster folding.
We note that the correct treatment of the total c.m. motion for the
$B_8 \alpha$ system is very important, since the c.m. correction
in the standard approach is of the order of $1/A \sim 1/4$.
As the result, the zero-momentum Wigner transform becomes
very much short-ranged and deep with the interaction range
about $R=1.2 A^{1/3} \sim 2~\hbox{fm}$.
The $\alpha$-particle density in \eq{fm4-12} is more compact
than the density including the c.m. motion,
$\rho(r)=4(2\nu/\pi)^{3/2}\,e^{-2\nu r^2}$, and the central
density is about $(4/3)^{3/2} \sim 1.5$ times larger.
The extremely large $N\alpha$ Wigner transform in Fig.\,\ref{fig1a}
is because of the factor 2 in \eq{a9}, and also because the other
nucleon-exchange term than the knock-on term and the effect of the
exchange normalization kernel are neglected. We will deal with this case
in a separate paper.

\begin{table}[b]
\caption{Values of the zero-momentum Wigner transform
at the origin $\br=0$, $V^C_W(0, 0)$ and $V^{LS}_C(0, 0)$,
and the Scheerbaum-like
factor, $\widetilde{S}_B$, in \protect\eq{fm4-14}.
The unit is in MeV for $V^C_W(0, 0)$ and $V^{LS}_C(0, 0)$,
and in MeV\,$\hbox{fm}^5$ for $\widetilde{S}_B$.
The models are fss2 and FSS, and $qtq$ and cont. imply 
the $QTQ$ and continuous choices for intermediate spectra,
respectively, used in the $G$-matrix calculations.
}
\vspace{2mm}
\label{table1}
\begin{center}
\renewcommand{\arraystretch}{1.2}
\setlength{\tabcolsep}{5mm}
\begin{tabular}{ccrrrr}
\hline
  & $B$ & \multicolumn{2}{c}{FSS} & \multicolumn{2}{c}{fss2} \\
  & & $qtq$ & cont & $qtq$ & cont \\
\hline
  & $N$       &  $-112.8$  &  $-121.1$  &  $-112.6$  &   $-120.0$  \\
$V^C_W(0, 0)$ &  $\Lambda$ &  $-26.01$  &  $-26.03$  &  $-28.09$
  &   $-29.31$  \\
  &  $\Sigma$ &   59.33    &   48.72    &    51.76   &    47.24    \\
  &  $\Xi$    &    7.97    &   10.73    &    20.83   &    20.38    \\
\hline
  &  $N$         &  $-65.60$  &  $-67.65$  &  $-63.89$  &  $-65.31$  \\
$V^{LS}_W(0, 0)$ &  $\Lambda$ &  $-14.56$  &  $-14.70$  &  $-17.89$
  &  $-18.91$  \\
  &  $\Sigma$    &  $-28.07$  &  $-23.45$  &  $-24.32$  &  $-22.92$  \\
  &  $\Xi$       &   18.59    &   25.55    &   7.71     &    8.59    \\
\hline
   &  $N$         &  $-51.00$  &  $-52.24$  &  $-52.15$  &  $-53.49$  \\
$\widetilde{S}_B$ &  $\Lambda$ &  $-5.32$   &  $-5.04$   &  $-13.22$
   &  $-13.96$  \\
   &  $\Sigma$    &  $-34.87$  &  $-30.36$  &  $-31.69$  &  $-30.89$  \\
   &  $\Xi$       &   20.15    &   26.30    &   6.77     &    7.62    \\
\hline
\end{tabular}
\end{center}
\end{table}

We list in Table \ref{table1} the values of the zero-momentum
Wigner transform at the origin $\br=0$, and the Scheerbaum-like
factor $\widetilde{S}_B$ in \eq{fm4-14} for all the possible
combinations of $G$-matrix calculations for the models, fss2 and FSS,
and for the $QTQ$ and continuous choices for intermediate
spectra. The corresponding values for the single-particle potentials,
$U_B(q_1)$, and the Scheerbaum factors, $S_B(q_1)$, at $q_1=0$ in
symmetric nuclear matter with $k_F=1.35~\hbox{fm}^{-1}$ are listed
in Table \ref{table2}. By comparing these results, we obtain the
following findings:
\begin{enumerate}
\item[1.] The $\Lambda \alpha$ central Wigner transforms are fairly
shallow in comparison with $U_\Lambda(0)$ in the symmetric nuclear matter
calculations. Namely, $|V^C_W(0,0)|$ is less than 30 MeV in all the
cases, while $|U_\Lambda(0)|$ is more than 40 MeV.
\item[2.] The $\Sigma \alpha$ and $\Xi \alpha$ central Wigner transforms
are repulsive, although $U_\Xi(0)$ is attractive.
In particular, $V^C_W(0,0)$ for $\Sigma \alpha$ is strongly repulsive,
which is due to the quark-model prediction of the repulsive $\Sigma N
(I=3/2)$ $\hbox{}^3S_1$ interaction. These characters are the
result of strong isospin dependence of the $\Sigma N$ and $\Xi N$
interactions, which is discussed in the next subsection.
\item[3.] The $LS$ component, $V^{LS}_W(0,0)$, for the $\Lambda \alpha$
interaction is by no means extremely small, in comparison with that
for the $\Sigma \alpha$ interaction. The ratio is only 70 -- 80 \%
for fss2 and 50 -- 60 \% for FSS. On the other hand, $\widetilde{S}_B$
factors reflect the characteristics of $S_B(0)$; namely,
$\widetilde{S}_B$ is about 20 -- 30 \% larger than $S_B(0)$, which is about
equal to the enhancement factor $\frac{1+4\xi}{4\xi} \sim 1.25$ -- 1.35 in
Eqs.\,(\ref{fm4-14}) and (\ref{fm4-18}).
\end{enumerate}

\begin{table}[t]
\caption{The depth of the single-particle potentials,
$U_B(q_1)$, and the Scheerbaum factors, $S_B(q_1)$, at  $q_1=0$ obtained
by the angular-averaged $G$-matrix calculations of the quark-model
potentials in symmetric nuclear matter.
The Fermi-momentum $k_F=1.35~\hbox{fm}^{-1}$ is used.
The models are fss2 and FSS, and $qtq$ and cont. imply 
the $QTQ$ and continuous choices for intermediate spectra,
respectively.
}
\vspace{2mm}
\label{table2}
\begin{center}
\renewcommand{\arraystretch}{1.2}
\setlength{\tabcolsep}{5mm}
\begin{tabular}{ccrrrr}
\hline
  & $B$ & \multicolumn{2}{c}{FSS} & \multicolumn{2}{c}{fss2} \\
  & & $qtq$ & cont & $qtq$ & cont \\
\hline
  &  $N$      &  $-79.8$  &  $-89.3$  &  $-80.6$  &   $-88.9$  \\
$U_B(0)$ & $\Lambda$ &  $-42.9$  &  $-46.3$  &  $-44.8$  &   $-48.4$  \\
  &  $\Sigma$ &   16.1    &   17.3    &    9.5    &    7.3    \\
  &  $\Xi$    &  $-14.9$  & $-20.8$   &   $-5.3$  &   $-8.0$  \\
\hline
  &  $N$ &   $-40.3$  &  $-41.4$  &  $-41.3$  &  $-42.4$  \\
$S_B(0)$ & $\Lambda$  &   $-3.9$  &   $-3.6$  &  $-10.0$  &  $-10.6$  \\
  &  $\Sigma$    &  $-27.4$  &  $-22.6$  &  $-24.1$  &  $-23.2$  \\
  &  $\Xi$       &   14.6    &   21.4    &   4.6     &    5.8    \\
\hline
\end{tabular}
\end{center}
\end{table}

\begin{figure}[t]
\begin{center}
\begin{minipage}[h]{0.49\textwidth}
\includegraphics[width=\textwidth]{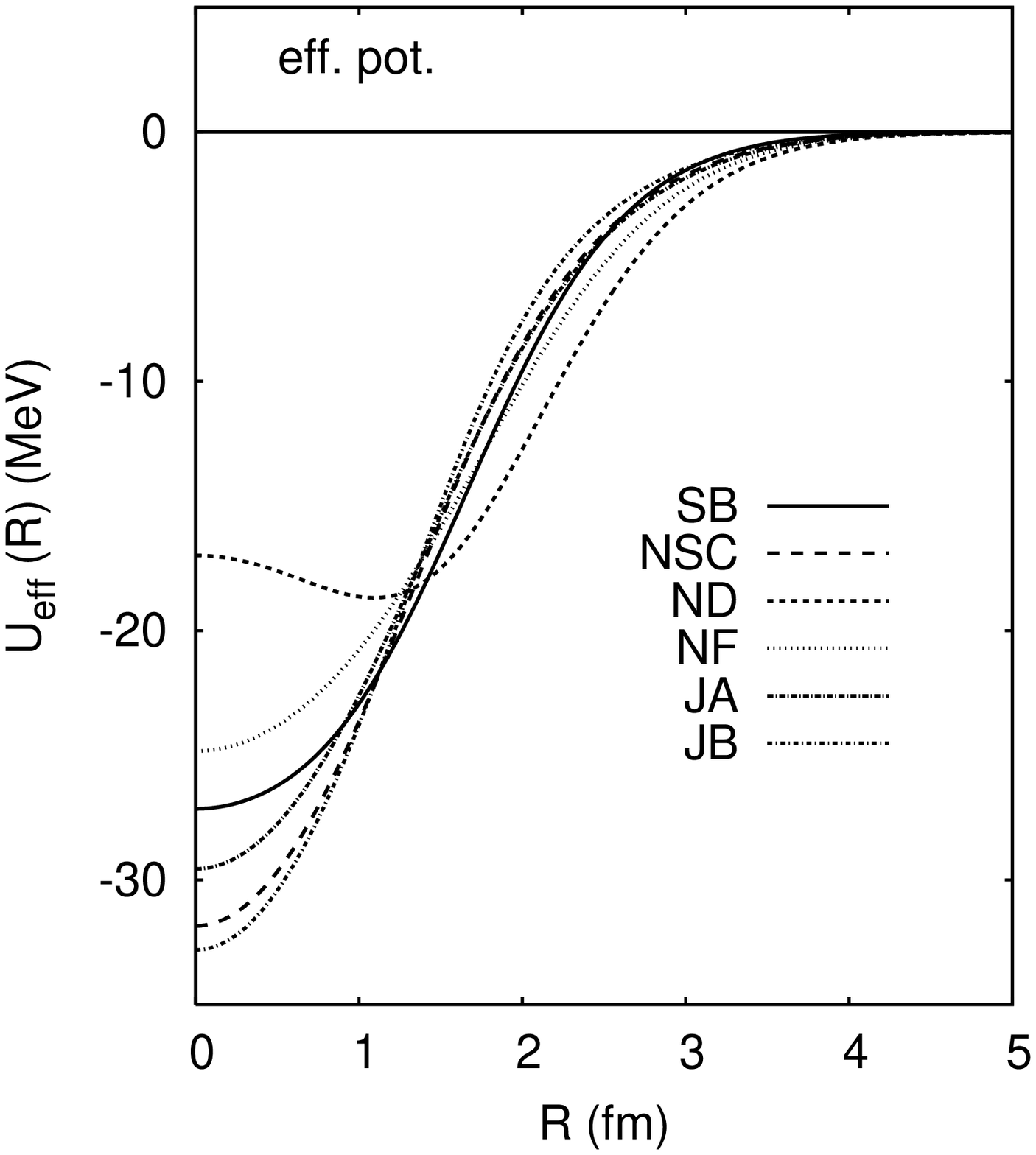}
\caption{Solutions of the transcendental equation \protect\eq{res1-3},
obtained from the $\Lambda \alpha$ Wigner transform
for the various effective $\Lambda N$
potentials. SB is the Sparenberg-Baye potential, Eqs.\,(\ref{res1-1})
and (\ref{res1-2}), NSC, ND, NF are the
simulated versions of the Nijmegen potentials, and JA, JB are
those of the J{\"u}lich potentials.
The energy $E=-3.12$ MeV is assumed.
The $(0s)^4$ shell-model wave function with the h.o. size
parameter $\nu=0.257~\hbox{fm}^{-2}$ is used
for the $\alpha$-cluster.
}
\label{fig3a}
\end{minipage}
\hfill
\begin{minipage}[h]{0.49\textwidth}
\includegraphics[width=\textwidth]{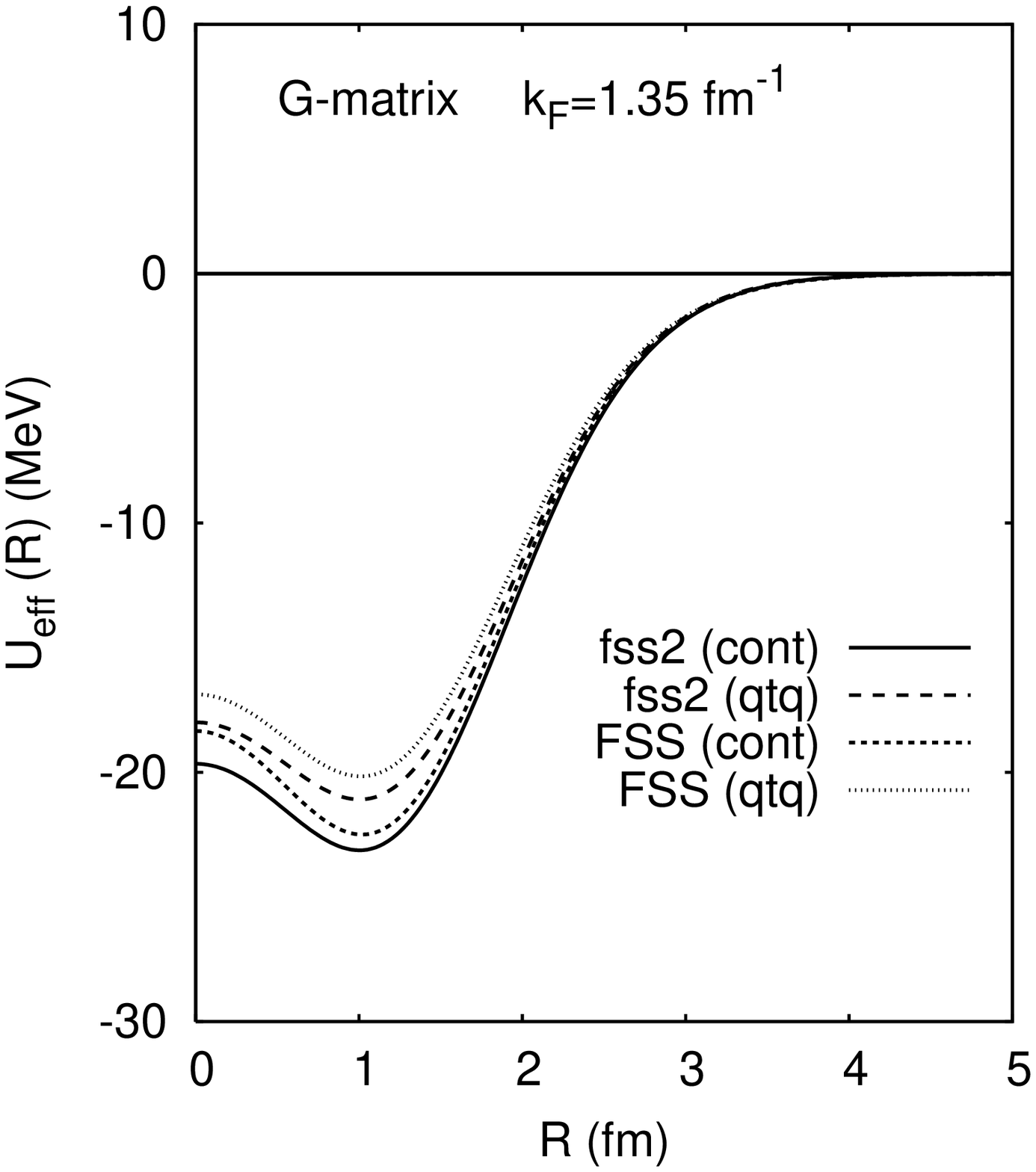}
\caption{
The same as Fig.~\protect\ref{fig3a}, but for
the Wigner transform, $V^C_W(\br,\bq)$ in \protect\eq{fm4-4},
calculated from the quark-model $G$-matrix interactions.
The $\Lambda \alpha$ Born kernels are
calculated from the quark-model $G$-matrix
$B_8 B_8$ interactions by fss2 and FSS. The Fermi momentum used
in the $G$-matrix calculation is $k_F=1.35~\hbox{fm}^{-1}$,
and $qtq$ stands for the $QTQ$ prescription
and cont for the continuous choice for intermediate spectra.
}
\label{fig3b}
\end{minipage}
\end{center}
\end{figure}

We find that the zero-momentum Wigner transform is not good enough
to define phase-shift equivalent local potentials, especially
for the $\Lambda \alpha$ interaction, In order to see this
clearly, we examined the Wigner transform for several effective
$\Lambda \alpha$ potentials, which can be easily derived from
the $\Lambda \alpha$ Born kernels
given in Appendix B of Ref.\,\cite{2al}.
The effective $\Lambda N$ potentials we examined are
the Sparenberg-Baye potential (SB potential) \cite{2al} given by
\begin{equation}
v_{\Lambda N}=\left[\,v(\hbox{}^1E) \frac{1-P_\sigma}{2}
+v(\hbox{}^3E) \frac{1+P_\sigma}{2}\,\right]
\,\left[\,\frac{u}{2}+\frac{2-u}{2}P_r\,\right]\ ,
\label{res1-1}
\end{equation}
with
\begin{eqnarray}
v(\hbox{}^1E) & = & -128.0~\exp(-0.8908~r^2)
+1015~\exp(-5.383~r^2)\ ,\nonumber \\
v(\hbox{}^3E) & = & -56.31\,f\,\exp(-0.7517~r^2)
+1072~\exp(-13.74~r^2)\ ,
\label{res1-2}
\end{eqnarray}
and the $G$-matrix simulated $\Lambda N$ forces \cite{HI97}
generated from the various OBEP potentials, NS (Nijmegen soft-core model
NSC89), ND (hard-core model D) , NF (hard-core model F),
JA (J{\"u}lich model A), and JB (model B).
By choosing the parameters, $u=0.946 87$ and $f=0.8923$ in 
Eqs.\,(\ref{res1-1}) and (\ref{res1-2}), we can correctly reproduce
the $\Lambda$ separation energy in $\hbox{}^5_\Lambda \hbox{He}$:
$E^{\rm exp}_B(\hbox{}^5_\Lambda \hbox{He})=-3.12 \pm 0.02~\hbox{MeV}$.
The strength of the short-range repulsive term (the third component)
of the NS - JB potentials are slightly modified from the original
values, in order to reproduce this value. (See Ref.\,\cite{2al}.)
The phase-shift equivalent local potential
in the semi-classical WKB-RGM approximation \cite{HO80,SHT84,NA95}
is calculated by solving the transcendental equation
\begin{equation}
U_{\rm eff}(R)
=G^{W}\left( R\,,~\sqrt{(2\mu_{\Lambda \alpha}/\hbar^2) \left[
E-U_{\rm eff}(R) \right]}\right)\ \ ,
\label{res1-3}
\end{equation}
for some specific energies $E$, where $G^W(R,\,q)$ is
assigned to $V^C_W(\br, \bq)$ in \eq{fm4-4}
with $R=r=|\br|$ and $q=|\bq|$.\footnote{For the
effective $\Lambda N$ forces,
the approximate formula \eq{fm4-4} gives an exact result,
since $\CV^C(\bk, \bq)$ is $(\bk \cdot \bq)$-independent.}
We here study only the $S$ wave, by neglecting
the usual semi-classical centrifugal term $\hbar^2(\ell+1/2)^2/
2\mu_{\Lambda \alpha}$.
The centrifugal potentials are included only in the 
Schr{\"o}dinger equation. This is a plausible approximation,
since the $LS$ term of the $\Lambda \alpha$ interaction is very small.
The obtained $\Lambda \alpha$ effective local potentials
with $E=-3.12~\hbox{MeV}$ are plotted in Fig.\,\ref{fig3a}
for SB - JA potentials.
The depth of $U_{\rm eff}$ is tabulated
in Table \ref{table3}, together with the $q^2$ value at $R=0$,
determined self-consistently. The bound-state energy of this
effective local potential, $E_B$, is calculated by solving the
Schr{\"o}dinger equation
\begin{equation}
\left[ -\frac{\hbar^2}{2\mu_{\Lambda \alpha}}
\frac{\partial^2}{(\partial R)^2}+U_{\rm eff}(R)\right]
\Psi(R)=E_B\,\Psi(R)\ \ .
\label{res1-4}
\end{equation}
We find that the bound-state energies of SB - JA potentials are
too small, compared with the exact value $E_B({\rm exact})$, which is obtained
by solving the Lippmann-Schwinger equation using the $\Lambda \alpha$
Born kernels.
Namely, $E_B$ from $U_{\rm eff}(R)$ is from 1.3 MeV to 1.7 MeV too small
in magnitude, except for the rather moderate difference 0.74 MeV in ND.
Figure \ref{fig3a} shows that this difference
is related with the interaction range of $U_{\rm eff}(R)$;
i.e., the range of ND is long while the others are
short. This poor result of the WKB-RGM approximation
for the $\Lambda \alpha$ interaction is probably related to the
very strong nonlocality (or momentum dependence) originating from
the $P_r$ term in \eq{res1-1}.
In order to see this, we artificially changed the Majorana exchange
mixture parameter $u$ in \eq{res1-1} and compared $E_B$ obtained
by the Wigner transform technique and by the exact method
using the $\Lambda \alpha$ Born kernel.
Table 4 shows this comparison. The case $u=2$ corresponds to pure
Wigner-type $\Lambda \alpha$ interaction, which gives a local
$\Lambda \alpha$ potential and complete agreement between the
two methods. Once we decrease the $u$ value and introduce
the Majorana component, the Wigner transform technique loses
the attractive effect of nonlocality very much and eventually
reaches at a very weak effective local potential with no bound state
before $u=0$ (the strength of the odd force =$-$(the strength
of the even force)). Our case is just in the middle of these two extremes,
which corresponds to the approximate Serber-type interaction with
a weak odd force. On the other hand, the exact solution is
almost independent of the $u$ value, which implies that the
$\Lambda$-particle is bound to the $\alpha$-cluster in the
almost $S$ wave.   
 
\begin{table}[t]
\caption{The depth of the effective local potential
$U_{\rm eff}(0)$, obtained by solving the transcendental equation
\protect\eq{res1-3} with $E=-3.12$ MeV. The $q^2$ value at $R=0$ and
a special value $V^C_W(0,0)$ of the $\Lambda \alpha$ Wigner transform
at $\br=0$ and $\bq=0$ are also given.
The eigenvalue $E_B$ is obtained by solving
the $S$-wave Schr{\"o}dinger equation \protect\eq{res1-4}
for $U_{\rm eff}(R)$.
The heading $E_B$ (exact) indicates the exact eigenvalue,
calculated from the $\Lambda \alpha$ Born kernel.
The quark-model $G$-matrix interactions are fss2 and FSS
both in the $QTQ$ ($qtq$) and continuous (cont) prescriptions
for intermediate spectra.  
The depth of the $LS$ potential $U^{LS}_{\rm eff}(0)$ is calculated from the
$q^2$ value determined by using only the central force.
}
\vspace{2mm}
\label{table3}
\begin{center}
\renewcommand{\arraystretch}{1.4}
\setlength{\tabcolsep}{3mm}
\begin{tabular}{ccccccc}
\hline
model & $V^C_W(0,0)$ & $q^2$ & $U_{\rm eff}(0)$ &
$U_{\rm eff}^{LS}(0)$ & $E_B$ & $E_B$ (exact) \\
      &  (MeV) & (fm$^{-2}$) & (MeV)  & (MeV) & (MeV) \\
\hline
  SB  & $-58.86$ & 1.062 & $-27.15$ & & $-1.86$  &  $-3.12$ \\
  NS  & $-71.00$ & 1.269 & $-31.85$ & & $-1.74$  &  $-3.12$ \\
  ND  & $-32.08$ & 0.613 & $-16.98$ & & $-2.38$  &  $-3.12$ \\
  NF  & $-61.91$ & 0.959 & $-24.83$ & & $-1.78$  &  $-3.12$ \\
  JA  & $-74.35$ & 1.168 & $-29.56$ & & $-1.60$  &  $-3.12$ \\
  JB  & $-83.71$ & 1.312 & $-32.81$ & & $-1.43$  &  $-3.12$ \\
\hline
fss2 (cont)  & $-29.53$ & 0.731 & $-19.66$ & $-17.22$ & $-2.94$  &  $-3.62$ \\
fss2 ($qtq$) & $-28.04$ & 0.657 & $-17.99$ & $-16.24$ & $-2.15$  &  $-2.75$ \\
FSS (cont)   & $-26.35$ & 0.673 & $-18.34$ & $-11.50$ & $-2.62$  &  $-3.18$ \\
FSS ($qtq$)  & $-25.95$ & 0.607 & $-16.87$ & $-12.30$ & $-1.78$  &  $-2.29$ \\
\hline
\end{tabular}
\end{center}
\end{table}

\begin{table}[htb]
\caption{
Comparison of the $S$-wave bound-state energies calculated
from the Wigner transform technique ($E_B$) and
the Lippmann-Schwinger approach of the $\Lambda \alpha$
Born kernel ($E_B({\rm exact})$), when the
Majorana exchange parameter $u$ in \eq{res1-1} is
changed from $u=2$ (pure Wigner) to $u=0$ (pure Majorana).
}
\vspace{2mm}
\label{table4}
\begin{center}
\renewcommand{\arraystretch}{1.4}
\setlength{\tabcolsep}{4mm}
\begin{tabular}{crr}
\hline
$u$ & $E_B$ & $E_B$ (exact) \\
    &  (MeV)   & (MeV)  \\
\hline
  2  & $-3.219$  & $-3.219$  \\
  1  & $-1.864$  & $-3.120$  \\
 0.5 & $-1.272$  & $-3.079$  \\
  0  & unbound   & $-3.041$ \\
\hline
\end{tabular}
\end{center}
\end{table}

\begin{figure}[t]
\begin{center}
\begin{minipage}[h]{0.49\textwidth}
\includegraphics[width=\textwidth]{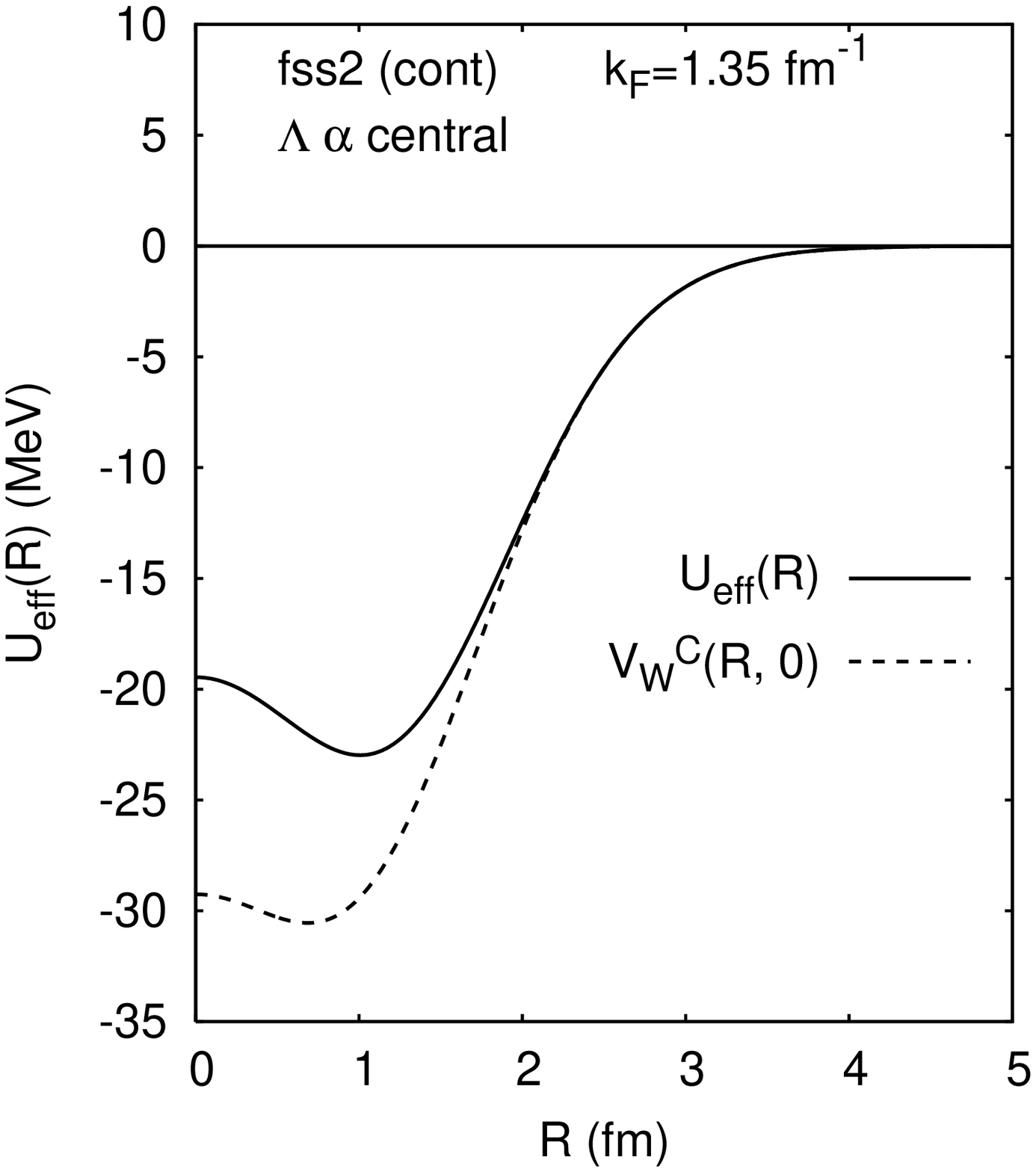}
\caption{The central component of the zero-momentum Wigner transform,
$V^C_W(R, 0)$, and the solution of the transcendental equation,
$U_{\rm eff}(R)$, for the quark-model $G$-matrix interaction
fss2. The continuous choice for intermediate spectra 
is used for the $G$-matrix calculation.
}
\label{fig5a}
\end{minipage}
\hfill
\begin{minipage}[h]{0.49\textwidth}
\vspace{-18mm}
\includegraphics[width=\textwidth]{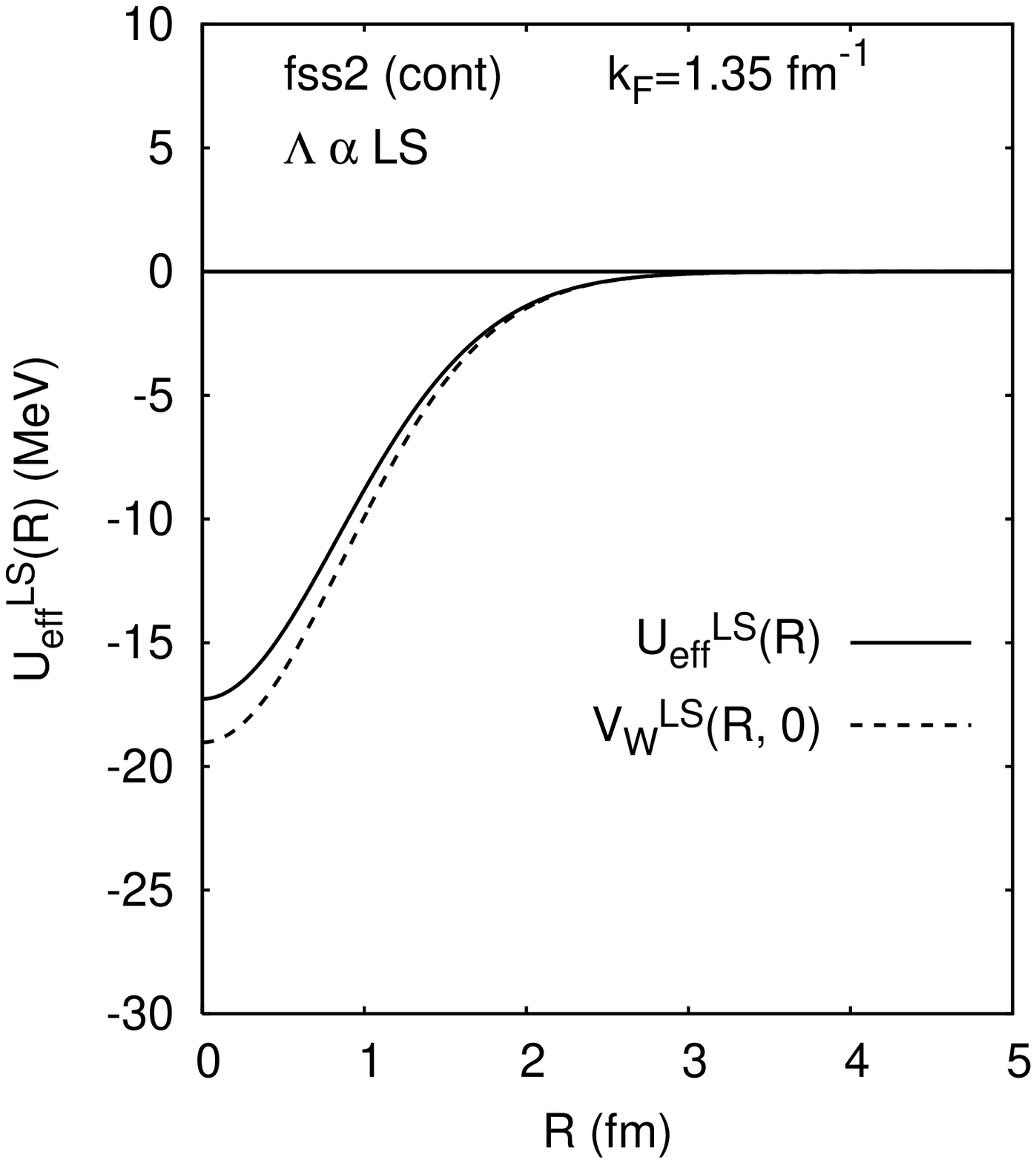}
\caption{
The same as Fig.~\protect\ref{fig5a}, but for
the $LS$ component.
The local momentum $q$ is determined self-consistently
by using only the central components as in Fig.~\protect\ref{fig5a}.
}
\label{fig5b}
\end{minipage}
\end{center}
\end{figure}

\begin{figure}[b]
\begin{center}
\begin{minipage}[h]{0.49\textwidth}
\vspace{-15mm}
\includegraphics[width=\textwidth]{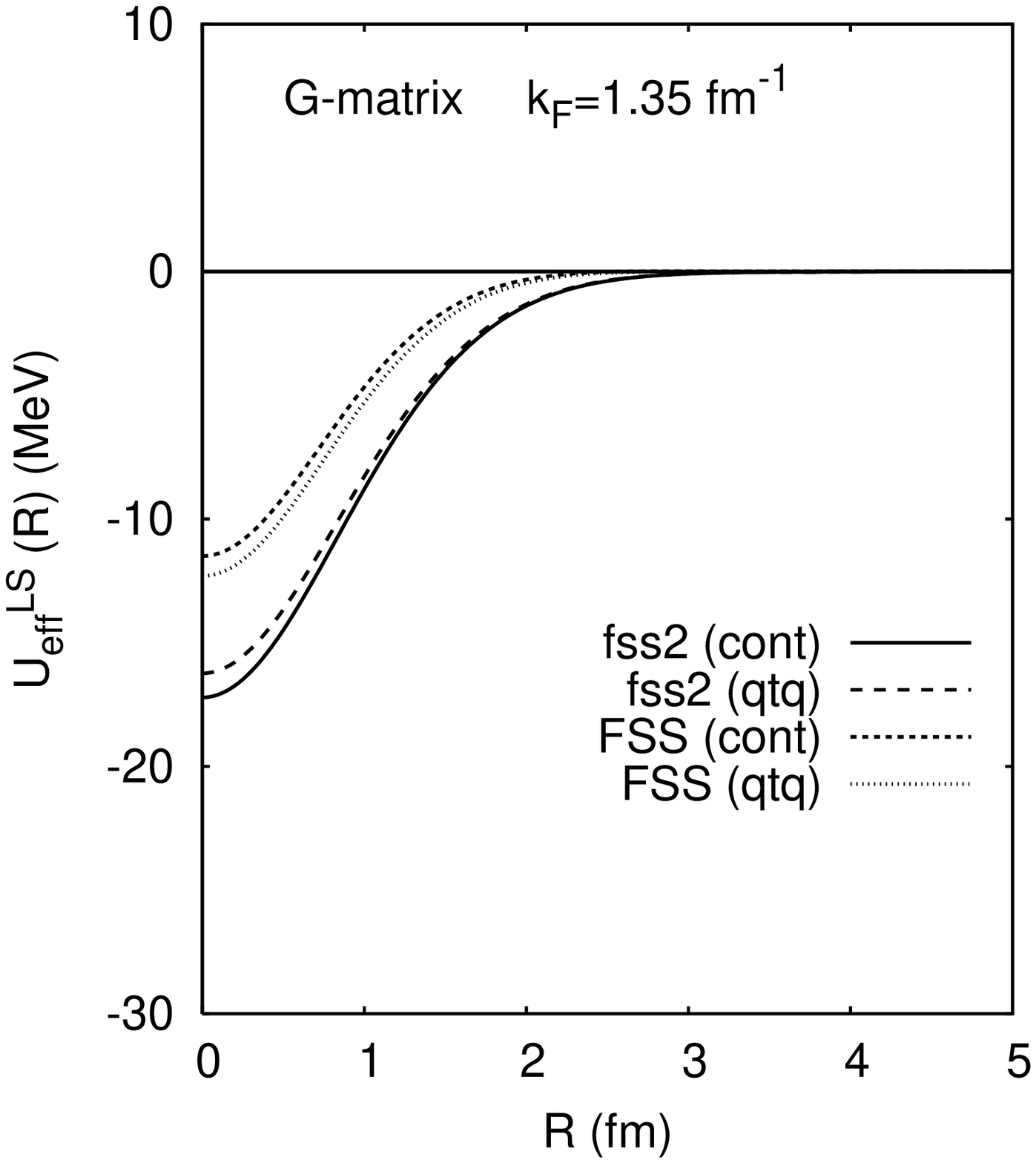}
\caption{
The same as Fig.\,\protect\ref{fig3b}
but for the $LS$ components. The local momentum
is determined only by the central force.
}
\label{fig7a}
\end{minipage}
\hfill
\begin{minipage}[h]{0.49\textwidth}
\includegraphics[width=\textwidth]{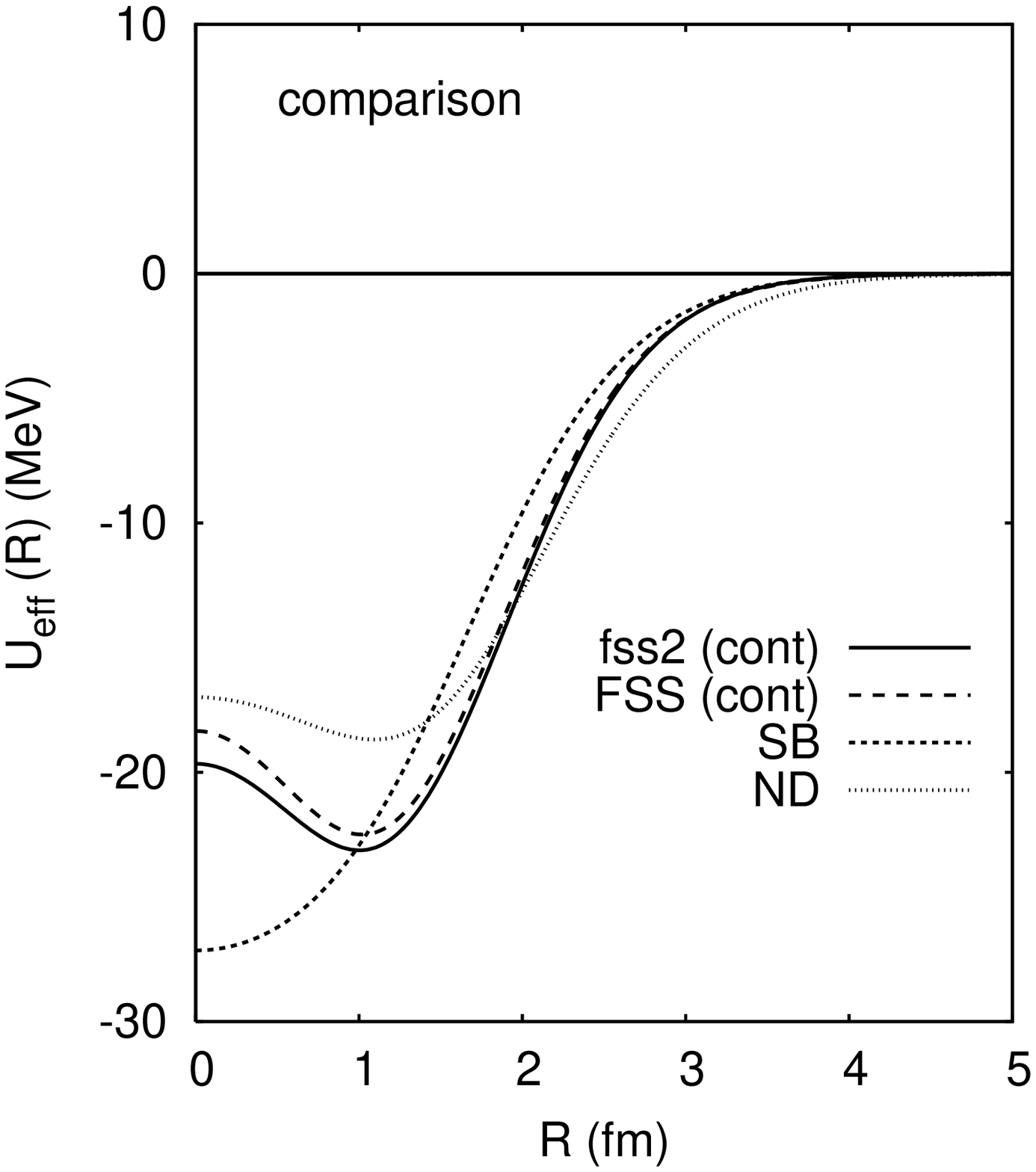}
\caption{
Comparison of effective local potentials
predicted by the quark-model $G$-matrix interactions
fss2, FSS (with the continuous choice for intermediate spectra)
and by some of the effective $\Lambda N$ forces, SB and ND
potentials.
}
\label{fig7b}
\end{minipage}
\end{center}
\end{figure}

The situation is almost the same even with the Wigner transform
approach of the quark-model $G$-matrix interaction. We show in
Fig.\,\ref{fig3b} solutions of the transcendental equations
obtained from the Wigner transform of $\Lambda \alpha$ Born kernels.
Here we also assumed $E=-3.12$ MeV. We find that different 
prescriptions for the $G$-matrix calculations, the $QTQ$ ($qtq$) or
the continuous (cont) choice for intermediate spectra, give essentially
same result with a slightly smaller attraction for the $QTQ$.
The model FSS gives a little weaker attraction than fss2.
The bound-state energies listed in Table \ref{table3} show
that the Schr{\"o}dinger equation of $U_{\rm eff}(R)$ gives
smaller energies by 0.5 $\sim$ 0.7 MeV than the exact method
using the $\Lambda \alpha$ Born kernel.
In fact, the bound-state energy for $\hbox{}^5_\Lambda \hbox{He}$,
obtained by solving the Lippmann-Schwinger equation is
$-3.62$ MeV for fss2 (cont) and $-3.18$ MeV for FSS (cont).
These values are by no means too large, compared with the experimental
value $E^{\rm exp}_B(\hbox{}^5_\Lambda \hbox{He})=-3.12 \pm 0.02~\hbox{MeV}$.
Table \ref{table3} also shows that the attraction is reduced
when the zero-momentum Wigner transform is converted to the effective
local potential. This is depicted in Fig.\,\ref{fig5a} for a typical
example of the fss2 prediction with the continuous prescription.
The small difference of $V^C_W(0, 0)$ values in Table \ref{table3}
from those in Table \ref{table1} is because $r=0.1~\hbox{fm}$
and $q=0.013~\hbox{fm}^{-1}$ are actually used in Table \ref{table1},
instead of $r=q=0$, and also because a spline interpolation
for $q$ is applied to $\CV^C_0(k, q)$ in \eq{fm4-4},
to facilitate the Wigner transform for an arbitrary $q$.
The $LS$ component $U_{\rm eff}^{LS}(R)$ depicted in Figs.\,\ref{fig5b}
and \ref{fig7a}, and the depth $U_{\rm eff}^{LS}(0)$ in Table \ref{table3}
are calculated from the $LS$ Wigner transform in \eq{fm4-4} by
using local momentum $q$ determined self-consistently for each $R$,
with respect to the central component.
First we find in Fig.\,\ref{fig5b} that the modification of the
zero-momentum $LS$ Wigner transform to the effective local potential is
comparatively small. Secondly, Fig.\,\ref{fig7a} shows that
the model FSS gives a shallower $LS$ potential
than model fss2, which is a result of the strong cancellation
between the ordinary $LS$ and the antisymmetric $LS$ ($LS^{(-)}$) forces
in the model FSS.
In Fig.\,\ref{fig7b}, we compare the
central components of the effective local potentials obtained
by the quark-model $G$-matrix interaction and by some of the effective
$\Lambda N$ forces. 
We find that the $G$-matrix prediction is rather long-ranged
and situated in the middle of ND and SB predictions.
The shape of our $\Lambda \alpha$ central potential is rather similar
to one of the phenomenological potentials in Ref.\,\cite{DA82}
(B of Fig.\,5), and to the effective potential derived from
a five-particle microscopic wave function with hard core correlations
in Ref.\,\cite{NA84} (the dashed curve of Fig.\,6).
It is convenient to parametrize the obtained $\Lambda \alpha$ potentials
in simple Gaussian functions. For example, the effective local potential
at $E=-3.12$ MeV, predicted by fss2 (cont) in Fig.\,\ref{fig7b},
is expressed as
\begin{equation}
U(R)=-(19.42+19.50\,R^2)\,e^{-0.5145\,R^2}\ \ (\hbox{MeV})\ \ ,
\label{res1-5}
\end{equation}
with the bound-state energy $E_B=-2.95$ MeV (which corresponds
to $-2.94$ MeV in Table \ref{table3}).

\begin{table}[t]
\caption{
The depths of the effective local potentials, $U_{\rm eff}(0)$, and
the $S$-wave phase shifts, $\delta^W_0(E)$, obtained by
solving the Schr{\"o}dinger equation.
The exact phase shift, $\delta_0(E)$ (exact), is also
shown, which is obtained by solving the Lippmann-Schwinger
equation from the Born kernel.
}
\vspace{2mm}
\label{table5}
\begin{center}
\renewcommand{\arraystretch}{1.4}
\setlength{\tabcolsep}{4mm}
\begin{tabular}{rrrrr}
\hline
model & $E_{\rm c.m.}$ & $U_{\rm eff}(0)$ & $\delta^W_0(E)$
      & $\delta_0(E)$ (exact) \\
      &  (MeV)   & (MeV)  & (deg) & (deg) \\
\hline
      &    10    & $-20.74$  & 62.74  & 72.20 \\
  SB  &    30    & $-11.90$  & 29.01  & 36.28 \\
      &    50    & $-4.26$   & 15.37  & 20.69 \\
\hline
      &    10    & $-25.16$  & 61.67  & 72.88 \\
  NS  &    30    & $-15.58$  & 28.12  & 37.10 \\
      &    50    & $-6.74$   & 13.46  & 20.46 \\
\hline
      &    10    & $-10.69$  & 62.14  & 68.39 \\
  ND  &    30    & $-1.78$   & 26.89  & 32.90 \\
      &    50    & 6.29      & 12.83  & 17.94 \\
\hline
fss2  &    10    & $-15.33$  & 70.58  & 74.82 \\
(cont)&    30    & $-9.46$   & 36.24  & 40.54 \\
      &    50    & $-4.04$   & 22.47  & 25.75 \\
\hline
fss2  &    10    & $-13.18$  & 64.92  & 70.07 \\
(qtq) &    30    & $-6.63$   & 31.28  & 35.50 \\
      &    50    & $-1.09$   & 18.18  & 21.54 \\
\hline
FSS   &    10    & $-14.60$  & 69.70  & 73.49 \\
(cont)&    30    & $-9.79$   & 36.88  & 40.50 \\
      &    50    & $-4.96$   & 23.01  & 26.62 \\
\hline
FSS   &    10    & $-12.21$  & 62.64  & 67.88 \\
(qtq) &    30    & $-6.10$   & 30.25  & 34.30 \\
      &    50    & $-1.27$   & 18.14  & 21.20 \\
\hline
\end{tabular}
\end{center}
\end{table}

We list in Table \ref{table5} the depths of the effective
local potentials $U_{\rm eff}(0)$ for positive energies,
$E_{\rm c.m.}=10,~30$ and 50 MeV, and the $S$-wave phase
shifts $\delta^W_0(E)$ obtained by
solving the Schr{\"o}dinger equation.
The exact phase shift $\delta_0(E)$ (exact) is also
shown, which is obtained from the Born kernel.
Here we again find the effective local potentials are too
shallow, especially for the effective $\Lambda N$ forces.
The phase shift difference is about 5 - $11^\circ$ for
the effective $\Lambda N$ forces, while 3 - $5^\circ$ for
the quark-model $G$-matrix interactions.
This implies that we need a readjustment of the effective
local potentials of the order of 10 MeV, in order to reproduce
the correct magnitude of the phase shifts.

Finally, we will make a brief comment on the choice of the $k_F$ value
in the present framework.
If we choose a smaller $k_F$, the $\Lambda$ s.p. potential in the
symmetric nuclear matter becomes shallower
and the $\Lambda \alpha$ central interaction becomes more attractive.
The first feature reduces the magnitude of
the starting energy
in the $\Lambda N$ $G$-matrix, resulting in more
attractive $\Lambda N$ $G$-matrix.
Although this change is compensated by the smaller volume
in the phase-space integral to calculate the $\Lambda$ s.p. potential,
such a mechanism does not work in the present
calculation of the $\Lambda \alpha$ interaction.
Thus the $\Lambda \alpha$ central interaction after the $\alpha$-cluster
folding becomes more attractive for a smaller $k_F$.
The same situation is observed in Fig.\,4 of Ref.\,\cite{YA94}
for their $\Sigma (3N)$ potential.
For example, if we change $k_F=1.35~\hbox{fm}^{-1}$
to $1.20~\hbox{fm}^{-1}$ (which corresponds to the 70$\%$ of the
normal density), the depth of the $\Lambda$ s.p. potential, $U_\Lambda(0)$,
in Table \ref{table2} is reduced from $-48.4$ MeV to $-37.5$ MeV for
the model fss2 in the continuous choice. On the other hand,
the depth of the effective local potential, $U_{\rm eff}(0)$,
and $E_B({\rm exact})$ in Table \ref{table3}
changes from $-19.66$ MeV to $-21.74$ MeV and from $-3.62$ MeV to
$-4.54$ MeV, respectively.\footnote{For the $LS$ component,
the $U^{LS}_{\rm eff}(0)$ values in Table \protect\ref{table3}
change from $-17.22$ MeV to $-17.23$ MeV for fss2 and
from $-11.50$ MeV to $-9.95$ MeV for FSS, when $k_F=1.20~\hbox{fm}^{-1}$
is used in the continuous choice.}
This implies that the self-consistent mechanism of the
starting-energy dependence, which is not properly taken into account
in this paper, is in fact very important. 
This finding is in accord with the importance
of the Brueckner rearrangement effect discussed in Ref.\,\cite{KO03}.
Since the purpose of the present study is not to examine the change
of the $\alpha$-cluster, it would be safe to assume the standard
value $k_F=1.35~\hbox{fm}^{-1}$, in order to examine the qualitative
features of the $B_8 \alpha$ interaction.

%\clearpage

%\vspace{-3mm}

\subsection{$\Sigma \alpha$ and $\Xi \alpha$ interactions}

\vspace{-3mm}

\begin{figure}[b]
\begin{center}
\begin{minipage}[h]{0.49\textwidth}
\includegraphics[width=\textwidth]{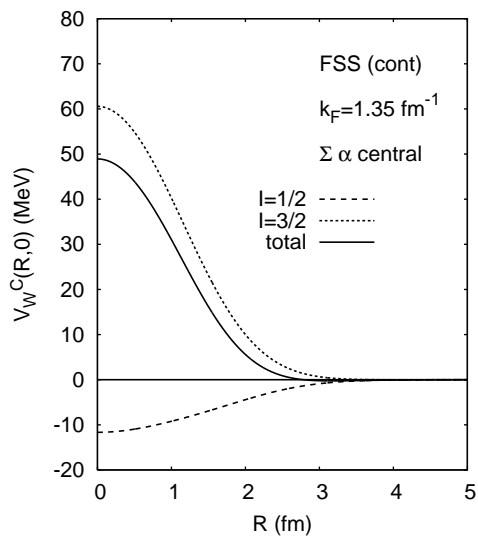}
\caption{
The central components of the zero-momentum Wigner
transform, ${V_W}^C(\br ,0)$ with $R=|\br|$,
for the $\Sigma \alpha$ Born kernel,
calculated from the quark-model $G$-matrix
$B_8 B_8$ interactions by FSS.
The Fermi momentum $k_F=1.35~\hbox{fm}^{-1}$ and the continuous choice
are used for the $G$-matrix calculation.
The h.o. size parameter $\nu=0.257~\hbox{fm}^{-2}$ is used
for the $\alpha$-cluster.
}
\label{fig9a}
\end{minipage}
\hfill
\begin{minipage}[h]{0.49\textwidth}
\includegraphics[width=\textwidth]{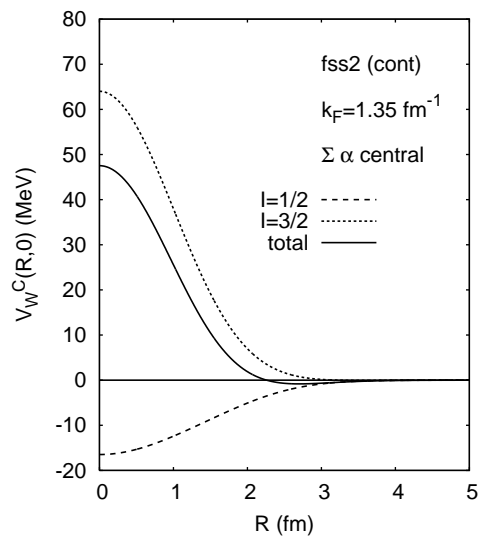}
\caption{
The same as Fig.\,\ref{fig9a}, but for fss2.
\ \vspace{37mm}
}
\label{fig9b}
\end{minipage}
\end{center}
\end{figure}

As seen from Fig.\,\ref{fig1a}, the $\Sigma \alpha$ and $\Xi \alpha$
interactions are repulsive. This does not mean that the $\Sigma$
and $\Xi$ potentials are always repulsive. The structure of
spin-isospin factors for the $\Sigma \alpha$ and $\Xi \alpha$ systems
is especially simple (see \eq{fm1-10}), which is due to the spin-isospin
saturated character of the $\alpha$-particle.  
It is, therefore, important to examine each isospin component
separately, in order to gain some insight to other possibilities
of unknown hypernuclei.
Here we discuss some qualitative features of these interactions,
based on the symmetry properties of the $B_8 B_8$ interactions
predicted by the quark-model interactions, FSS and fss2.
When the interaction is repulsive, the transcendental equation
\eq{res1-3} sometimes does not have its solution, since
the square of the local momentum, $q^2$, becomes negative.
Since the extension of the Wigner transform \eq{fm4-4} to
the negative $q^2$ is not easy numerically, we only discuss
the zero-momentum Wigner transform in this subsection.

\begin{figure}[t]
\begin{center}
\begin{minipage}[h]{0.49\textwidth}
\includegraphics[width=\textwidth]{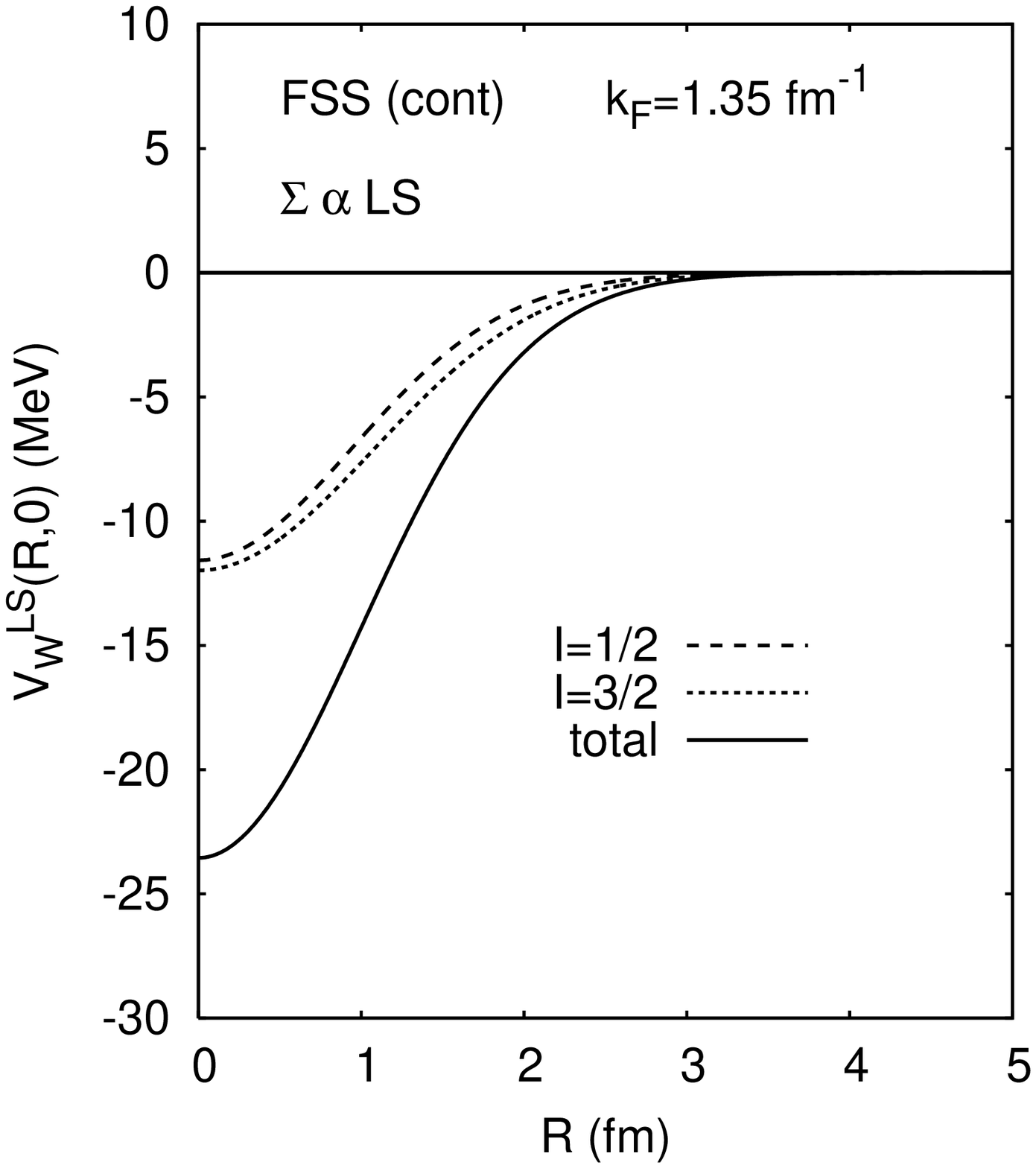}
\caption{
The same as Fig.\,\ref{fig9a}, but for the $LS$ components.
}
\label{fig11a}
\end{minipage}
\hfill
\begin{minipage}[h]{0.49\textwidth}
\includegraphics[width=\textwidth]{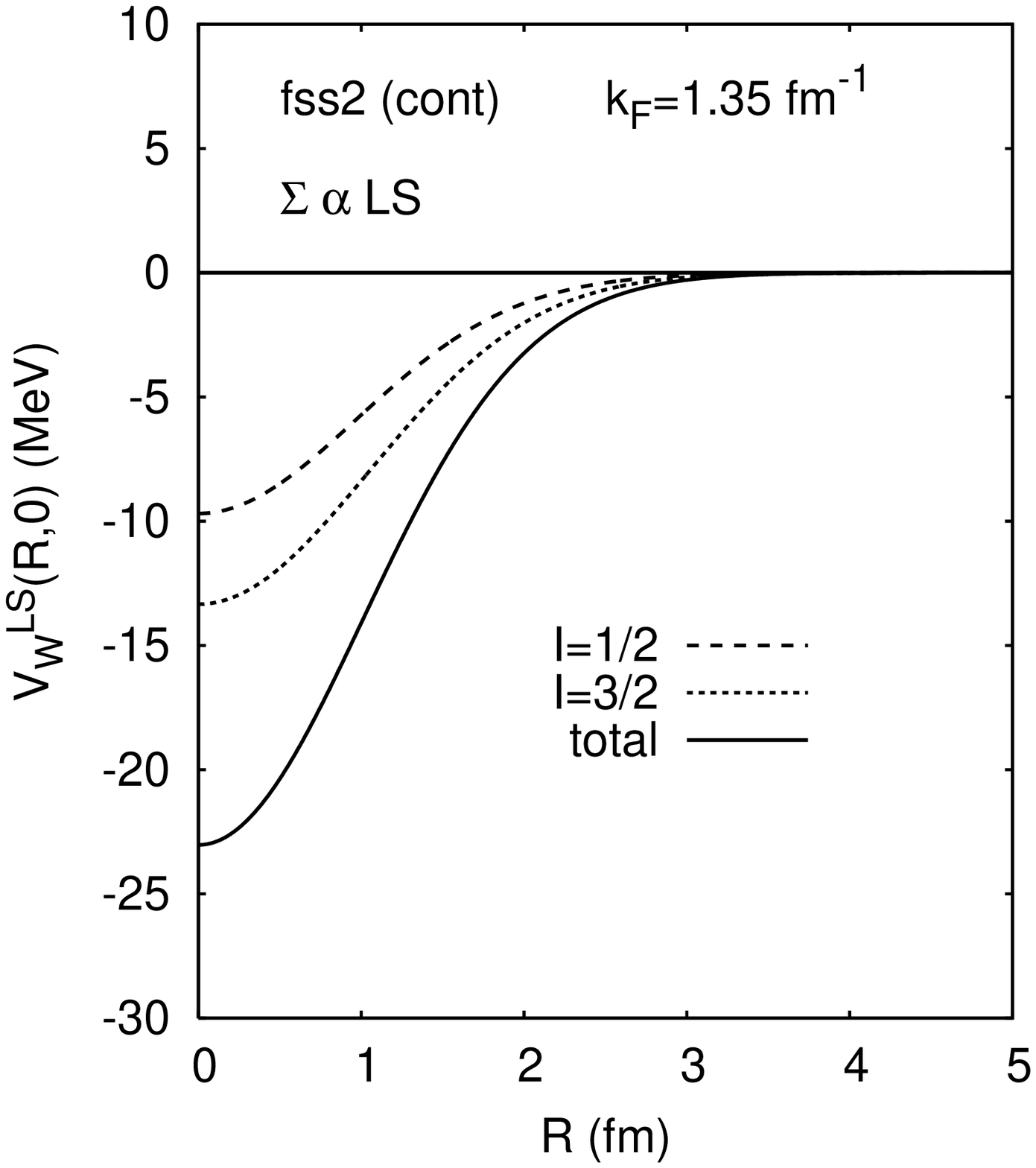}
\caption{
The same as Fig.\,\ref{fig11a}, but for fss2.
\vspace{1mm}
}
\label{fig11b}
\end{minipage}
\end{center}
\end{figure}

Figures \ref{fig9a} and \ref{fig9b} show the isospin components
with $I=1/2$ and $I=3/2$ for the $\Sigma \alpha$ central
interaction, predicted by FSS and fss2, respectively.
We find that both interactions have some amount of attraction
originating from the $\hbox{}^3S_1$ channel of the $I=1/2$
$\Sigma N$ interaction. This channel becomes attractive due to
the very strong $\Lambda N$--$\Sigma N$ coupling by the
one-pion exchange tensor force.
On the other hand, the $\hbox{}^3S_1$ state
of the $I=3/2$ channel is strongly repulsive due to the
Pauli principle at the quark level.  
We find from Figs. \ref{fig9a} and \ref{fig9b} that
this repulsion is so strong that almost no attraction from the $I=1/2$
channel remains in the $\Sigma \alpha$ interaction.

\begin{figure}[t]
\begin{center}
\begin{minipage}[h]{0.49\textwidth}
\includegraphics[width=\textwidth]{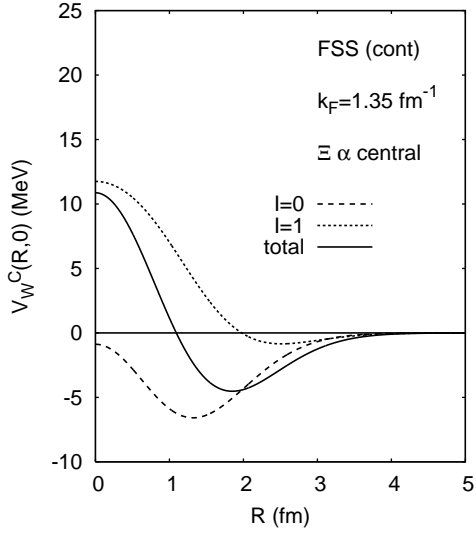}
\caption{
The same as Fig.\,\ref{fig9a}, but for the $\Xi \alpha$ interaction.
}
\label{fig13a}
\end{minipage}
\hfill
\begin{minipage}[h]{0.49\textwidth}
\includegraphics[width=\textwidth]{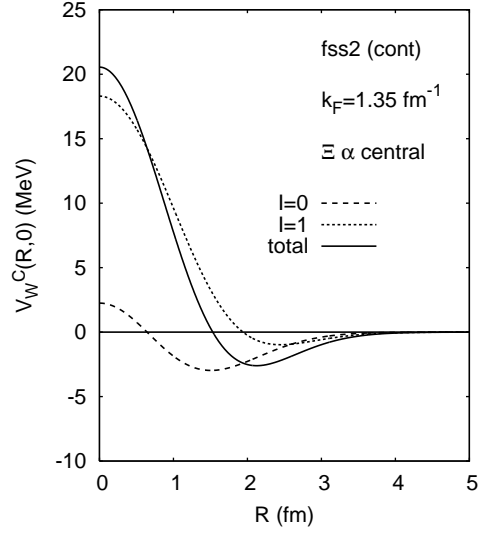}
\caption{
The same as Fig.\,\ref{fig13a}, but for fss2.
\vspace{1mm}
}
\label{fig13b}
\end{minipage}
\end{center}
\end{figure}

\begin{figure}[t]
\begin{center}
\begin{minipage}[h]{0.49\textwidth}
\includegraphics[width=\textwidth]{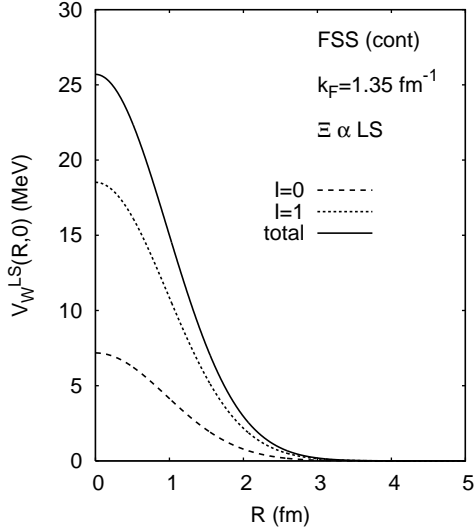}
\caption{
The same as Fig.\,\ref{fig13a}, but for the $LS$ components.
}
\label{fig15a}
\end{minipage}
\hfill
\begin{minipage}[h]{0.49\textwidth}
\includegraphics[width=\textwidth]{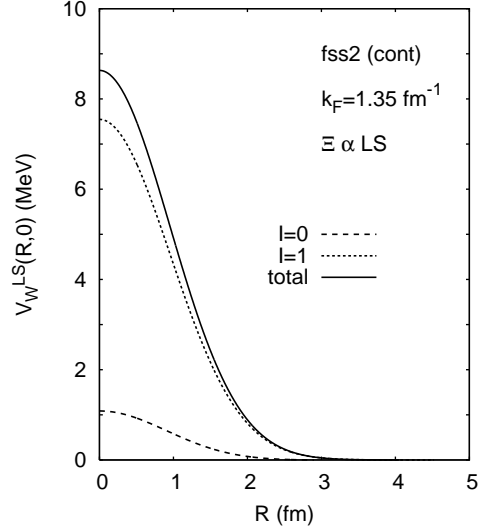}
\caption{
The same as Fig.\,\ref{fig15a}, but for fss2.
\vspace{1mm}
}
\label{fig15b}
\end{minipage}
\end{center}
\end{figure}

On the other hand, the two isospin channels, $I=1/2$ and 3/2,
yield fairly large $LS$ force for the $\Sigma \alpha$ interaction.
This can be seen in Figs.\,\ref{fig11a} and \ref{fig11b}.
In the isospin $I=3/2$ channel, $\hbox{}^3P_J$ states are classified
to the flavor symmetric channel with the $SU_3$ label (22).
It is, therefore, plausible that the same mechanism as the
$NN$ $I=1$ channel yields very strong $LS$ force. On the other hand,
a part of the $LS$ force from the isospin $I=1/2$ channel is due to
the flavor-exchange process between $(11)_a$ and $(11)_s$ $SU_3$
configurations. This process is accompanied with the spin flip
between $S=0$ and 1, and yields very strong $LS^{(-)}$ force.
The $LS$ and $LS^{(-)}$ forces reinforce each other in just opposite
way to the $\Lambda N$ interaction, and the resultant $\Sigma \alpha$
$LS$ force becomes almost 3/5 of the $N \alpha$ $LS$ force,
as seen from $\widetilde{S}_B$ in Table \ref{table1}.
This ratio is almost the same as that of the Scheerbaum
factors $S_B=S_B(0)$ in symmetric nuclear matter. (See Table \ref{table2}.)

Figures \ref{fig13a} and \ref{fig13b} show the isospin components
with $I=0$ and $I=1$ for the $\Xi \alpha$ central
interaction, predicted by FSS and fss2, respectively.
Here again we find the situation that the attractive
nature of the $I=0$ component is largely canceled by
the repulsion in the $I=1$ channel.
However, this cancellation is not strong especially in the model FSS,
and we have almost 5 MeV attraction around $R=2$ fm.
In fss2, the height of the central repulsion in the $I=1$ channel is
almost 20 MeV, and we can expect a few MeV attraction in the surface
region. These long-range attractions may have some influence to the
atomic orbit between $\Xi^-$ and $\alpha$.
The $\Xi \alpha$ potential rather similar (though more attractive) to
that by FSS is presented by Myint and Akaishi in Ref.\,\cite{MY94}.
It should be noted that the origin of the repulsion in the $I=1$ channel
is the Pauli forbidden state $(11)_s$ in the $\hbox{}^1S_0$ state
and the almost Pauli forbidden state (30) in the $\hbox{}^3S_1$ state.
However, the coupling with the $\Lambda \Sigma$ channel is
very important, which may cause the long-range attraction even in the
$I=1$ channel.

The $LS$ components of the $\Xi \alpha$ interaction is repulsive,
which is clearly seen in Figs.\,\ref{fig15a} ans \ref{fig15b}.
This $LS$ force is fairly strong, especially for FSS.
The magnitude is almost 1/3 of the $N \alpha$ $LS$ force with
the opposite sign. These features are very similar to the 
$LS$ force in symmetric nuclear matter, as seen in Table \ref{table2}.

%\clearpage
%\vspace{-3mm}

\section{Summary}

\vspace{-3mm}

The $SU_6$ quark-model baryon-baryon interaction (fss2, FSS)
\cite{PPNP,B8B8,fss2} is a
unified model which describes all the baryon-octet baryon-octet
($B_8 B_8$) interactions in a full coupled-channel formalism.
For the nucleon-nucleon ($NN$) and hyperon-nucleon ($YN$) interactions,
all the available experimental data are reasonably reproduced.
It is, therefore, interesting to study $B_8 \alpha$ interactions
in a microscopic framework under a simple assumption of
the $(0s)^4$ harmonic-oscillator shell-model wave function
for the $\alpha$-cluster.
In this study, we have used the result of $G$-matrix calculations
for symmetric nuclear matter, as an input for the two-body interactions
for the $\alpha$-cluster folding.
Since the resultant $B_8 \alpha$ interactions are rather insensitive
to the Fermi-momentum $k_F$ in the $G$-matrix calculations
(except for the $\Lambda N$ $LS$ force for FSS), we have assumed
$k_F=1.35~\hbox{fm}^{-1}$. The other $G$-matrix parameters,
the center-of-mass (c.m.) momentum $K$ of two interacting particles
and the starting energy $\omega$, are treated unambiguously
in the total c.m. frame of the $B_8 \alpha$ system.
This can be achieved by considering the transformation of the
matrix elements in the momentum representation from the initial
($\bq_i$) and final ($\bq_f$) momenta
to the momentum transfer ($\bk=\bq_f-\bq_i$) and
the local momentum $\bq=(\bq_f+\bq_i)/2$.
If one uses this transformation at the level of $G$-matrix,
the procedure of the $\alpha$-cluster folding becomes extremely
simple both for the central and $LS$ components.
The $B_8 \alpha$ interaction, $\CV^{\Omega} (\bk, \bq)$
with $\Omega=C,~LS$, represented by
$\bk$ and $\bq$ is then transformed back to the $B_8 \alpha$ Born
kernel, $V^{\Omega}(\bq_f, \bq_i)$, through the inverse transformation.
This procedure is also convenient to calculate the Wigner transform,
$V^{\Omega}_W (\br, \bq)$, which is simply a Fourier transform
of $\CV^{\Omega} (\bk, \bq)$. By solving the transcendental equation
for $V^{\Omega}_W (\br, \bq)$, we can obtain an energy-dependent
local potential of the $B_8 \alpha$ system in the WKB-RGM approximation
\cite{HO80,SHT84,NA95}.

In this paper, we have applied the present formalism to the 
$\Lambda \alpha$, $\Sigma \alpha$ and $\Xi \alpha$ systems.
Applications to the $N \alpha$ system will be discussed in a
separate paper, since this system involves
an extra nucleon-exchange term, in addition to the direct
and knock-on terms. In the $\Lambda \alpha$ system, the
WKB-RGM approximation is rather poor due to the very strong
momentum dependence of the exchange knock-on term.
This term appears even for the effective $\Lambda N$ force,
which is traced back to the strangeness exchange processes.
We find that the $\Lambda \alpha$ central potentials predicted by 
various quark-model $G$-matrices are very similar to each other,
irrespective of a specific model, fss2 or FSS, and 
the $QTQ$ or continuous choice for intermediate spectra.
At the bound-state energy, $E=-3.12$ MeV, they are long-range
local potentials with the wine-bottle shape,
having the depth less than 30 MeV.
They are very similar to the $\Lambda \alpha$ potential
obtained from the effective $\Lambda N$ potential ND \cite{HI97},
simulating the Nijmegen hard-core model D.
The $\Lambda \alpha$ bound-state energies are calculated
by solving the Lippmann-Schwinger equation
of the $\Lambda \alpha$ Born kernel.
These are $-3.62$ MeV for fss2 (cont) and $-3.18$ MeV for FSS (cont),
when $k_F=1.35~\hbox{fm}^{-1}$ is used.
It should be noted that the depth of the single-particle potential
for $\Lambda$ in symmetric nuclear matter is 48 MeV for fss2 (cont)
and 46 MeV for FSS (cont) \cite{PPNP}. 
The fact that the predicted $\Lambda \alpha$ bound-state energies
are by no means too large, in comparison with the experimental
value $E^{\rm exp}_B(\hbox{}^5_\Lambda \hbox{He})=-3.12 \pm 0.02~\hbox{MeV}$,
implies that the proper treatment
of the c.m. motion of the $\Lambda \alpha$ system is very
important. It is also important to note that, in the present
$G$-matrix approach, the $\Lambda N$--$\Sigma N$ coupling by
the very strong one-pion exchange tensor force is explicitly
treated, the lack of which is known to lead to the so-called
overbinding problem of the $\Lambda \alpha$ bound state.
The energy loss predicted by the $\alpha$-cluster rearrangement
effect \cite{KO03} through the starting-energy dependence
of the $G$-matrix needs further detailed analyses.
On the other hand, the $\Lambda \alpha$ $LS$ potentials are
rather model dependent. In the model fss2, the depth of the
$LS$ potential is $-16 \sim -17$ MeV, while in FSS about $-12$ MeV.
The interaction range of the FSS $LS$ potential is also 
very short, which leads to a small Scheerbaum-like factor
$\widetilde{S}_\Lambda$ for the $\Lambda \alpha$ $LS$ force.
We will show in a separate paper, the strength of the $LS$ force
is further reduced for smaller values of $k_F$, if FSS is used.
The very small spin-orbit splitting of the $\hbox{}^9_\Lambda
\hbox{Be}$ \cite{AK02,TA03} can be reproduced
in the $\alpha \alpha \Lambda$ Faddeev calculation,
using the $\alpha \alpha$ RGM kernel
and the $\Lambda \alpha$ $LS$ Born kernel predicted from
the FSS $G$-matrix with $k_F=1.25~\hbox{fm}^{-1}$.

Based on the reasonable reproduction of the $\Lambda \alpha$
interaction properties, we have examined the real parts
of the $\Sigma \alpha$ and $\Xi \alpha$ interactions in the
Wigner transform technique. Since these interaction are repulsive,
we have examined only the zero-momentum Wigner transform
as the first step. In the $\Sigma \alpha$ interaction,
the attractive effect from the isospin $I=1/2$ $\Sigma N$ channel
is completely cancelled out by the repulsion
from the $I=3/2$ $\Sigma N$ channel. The origin of this strong
repulsion is the $\Sigma N (I=3/2)$ $\hbox{}^3S_1$ channel,
which contains the almost Pauli-forbidden $SU_3$ (30)
component for the most compact $(0s)^6$ configuration.
On the other hand, the $\Xi \alpha$ interaction is
less repulsive because of the appreciable attraction originating
from the $I=0$ $\Xi N$ channels. The $\Xi \alpha$ zero-momentum
Wigner transform predicted by FSS yields about $-5$ MeV attraction
around $R=2$ fm, while the attraction of fss2 is about $-3$ MeV.
These long-range attractions may have some relevance to the
formation of atomic bound states for the $\Xi^- \alpha$ system.
As to the spin-orbit interaction, the two isospin channels
of the $\Sigma N$ interaction give fairly strong
attractive $\Sigma \alpha$ $LS$ forces, yielding almost
3/5 of the $N \alpha$ $LS$ force.
On the other hand, $\Xi \alpha$ $LS$ force is
repulsive and the magnitude is $1/8 \sim 1/2$ of the $N \alpha$ $LS$ force. 
We will show in the next paper, the present $N \alpha$ $LS$ force
is consistent with the observed $P$-wave splitting
of the $3/2^-$ and $1/2^-$ excited states of $\hbox{}^5\hbox{He}$.

It should be noted that the overall repulsive character of the
$\Sigma \alpha$ and $\Xi \alpha$ interactions is related to the
spin-isospin saturated character of the $\alpha$-cluster.
The strong isospin dependence of the $\Sigma N$ and $\Xi N$ interactions,
namely, repulsive for the $\Sigma N (I=3/2)$ and $\Xi N (I=1)$ channels
and attractive for the $\Sigma N (I=1/2)$ and $\Xi N (I=0)$ channels,
leads to a possibility of attractive features in some particular
spin-isospin channels for systems of $\Sigma$, $\Xi$ and the $s$-shell
clusters \cite{HA90,OK90,YA94}.
One of the examples of such systems is the isospin $I=1/2$
and spin $S=0$ state of $\hbox{}^4_\Sigma \hbox{He}$, in which
the strong repulsion of the $\Sigma N (I=3/2)$ $\hbox{}^3S_1$ channel
does not contribute \cite{YA92} and a quasi-bound state is in fact observed
experimentally \cite{HA89,NA98}.
Applications of the present approach to such systems are
under way.

\bigskip

\noindent
{\bf Acknowledgments}
%\begin{acknowledgments}

\bigskip

This work was supported by Grants-in-Aid for Scientific
Research (C) from the Japan Society for the Promotion
of Science (JSPS) (Grant Nos.~18540261 and 17540263).
%\end{acknowledgments}

\bigskip

\appendix

\section{Invariant $G$-matrix for the most general $B_8 B_8$ interaction}

\vspace{-3mm}

In this Appendix, we will define the invariant $G$-matrix for the
most general $B_8 B_8$ interaction.
The $B_8 B_8$ channels in the bra side ($\gamma$) and ket side ($\alpha$)
are specified by \cite{NA95}
\begin{eqnarray}
\gamma & = & \left[\frac{1}{2}(11)c_1\,\frac{1}{2}(11)c_2\right]
SS_zYII_z; \CP\ \ ,\nonumber \\
\alpha & = & \left[\frac{1}{2}(11)a_1\,\frac{1}{2}(11)a_2\right]
S^\prime {S_z}^\prime YII_z; {\CP}^\prime\ \ ,
\label{a1}
\end{eqnarray}
in the isospin basis. For example, the spin-flavor functions are given by
\begin{eqnarray}
& & \eta_\gamma=\chi_{SS_z}\,[B_1 B_2]^\CP_{II_z}\ \ ,\nonumber \\
& & [B_1 B_2]^\CP_{II_z}=\frac{1}{\sqrt{2(1+\delta_{c_1,c_2})}}
\left\{[B_1 B_2]_{II_z}+\CP (-1)^{I_1+I_2-I}[B_2 B_1]_{II_z}\right\}
\ ,
\label{a2}
\end{eqnarray}
where the isospin-coupled flavor wave functions, $[B_1 B_2]_{II_z}$, are  
lexicographically ordered and the first baryon is numbered always 1 and
the second 2 (i.e., $[B_2 B_1]_{II_z}=[B_2(1) B_1(2)]_{II_z}$).
The flavor symmetry basis, $[B_1 B_2]^\CP_{II_z}$ with $\CP=1$ and
$-1$, gives
\begin{eqnarray}
\left[B_1 B_2\right]_{II_z}
& = & \sqrt{\frac{1+\delta_{c_1, c_2}}{2}}
\sum_\CP \left[B_1 B_2\right]^\CP_{II_z}\ \ ,\nonumber \\
\left[B_3 B_4\right]_{II_z} & = & \sqrt{\frac{1+\delta_{a_1, a_2}}{2}}
\sum_{\CP^\prime} \left[B_3 B_4\right]^{\CP^\prime}_{II_z}\ \ .
\label{a3}
\end{eqnarray}
The matrix element of the $G$-matrix in the isospin basis
and its partial-wave decomposition are
given by \cite{LSRGM}
\begin{eqnarray}
G_{\gamma \alpha}(\bp, \bp^\prime; K, \omega)
& = & \langle \chi_{SS_z} [B_1 B_2]^\CP_{II_z} |
G(\bp, \bp^\prime; K, \omega) |
\chi_{S^\prime{S_z}^\prime} [B_3 B_4]^{\CP^\prime}_{II_z}\rangle
\nonumber \\
& = & \sum^\prime_{JM\ell \ell^\prime}
4\pi G^J_{\gamma S\ell,\alpha S^\prime{S_z}^\prime}(p, p^\prime;
K, \omega) \sum_m \langle \ell m S S_z | JM \rangle
\,Y_{\ell m}(\widehat{\bp}) \nonumber \\
& & \times \sum_{m^\prime}
\langle \ell^\prime m^\prime S^\prime {S_z}^\prime
| JM \rangle\,Y^*_{\ell^\prime m^\prime}(\widehat{\bp}^\prime)\ \ .
\label{a4}
\end{eqnarray}
Here the prime symbol on $\sum$ implies that the summation is only
for such quantum numbers that satisfy the generalized Pauli
principle:
\begin{equation}
(-1)^\ell (-1)^{1-S} \CP = (-1)^{\ell^\prime} (-1)^{1-S^\prime}
\CP^\prime = -1\ \ .
\label{a5}
\end{equation}
We multiply \eq{a4} with $|\chi_{SS_z}\rangle$ and 
$\langle \chi_{S^\prime {S_z}^\prime}|$ from the left- and
right-hand sides, respectively,
and take a sum over
$SS_z$ and $S^\prime {S_z}^\prime$. Then we find
\begin{eqnarray}
& & \langle [B_1 B_2]^\CP_{II_z} | G(\bp, \bp^\prime; K, \omega) |
[B_3 B_4]^{\CP^\prime}_{II_z}\rangle
\nonumber \\
& & =\sum^\prime_{JM\ell \ell^\prime S S^\prime}
4\pi\,G^J_{\gamma S\ell,\alpha S^\prime{S_z}^\prime}(p, p^\prime;
K, \omega)\,\CY_{(\ell S)JM}(\widehat{\bp}; {\rm spin})
\,\CY^*_{(\ell^\prime S^\prime)JM}(\widehat{\bp}^\prime; {\rm spin})
\ \ ,\nonumber \\
\label{a6}
\end{eqnarray}
where $\CY_{(\ell S)JM}(\widehat{\bp}; {\rm spin})
=[Y_\ell(\widehat{\bp})\chi_S]_{JM}$ is the angular-spin
function for the two-baryon system.

Let us denote the two-baryon channels
with $c=(c_1, c_2)$, $a=(a_1, a_2)$, and consider
\begin{eqnarray}
& & G_{ca}(\bp, \bp^\prime; K, \omega)
\nonumber \\
& & \equiv \langle [B_1 B_2]_{II_z} | G(\bp, \bp^\prime; K, \omega)
-G(\bp, -\bp^\prime; K, \omega) P_\sigma P_F | [B_3 B_4]_{II_z} \rangle
\ \ .
\label{a7}
\end{eqnarray}
We can rewrite this by using Eqs.\,(\ref{a3}) and (\ref{a6}):
\begin{eqnarray}
& & G_{ca}(\bp, \bp^\prime; K, \omega) \nonumber \\
& & =\frac{1}{2}\sqrt{(1+\delta_{c_1,c_2})(1+\delta_{a_1,a_2})}
\sum_{\CP, \CP^\prime} \left\{
\langle [B_1 B_2]^\CP_{II_z} | G(\bp, \bp^\prime; K, \omega) |
[B_3 B_4]^{\CP^\prime}_{II_z}\rangle
\right. \nonumber \\
& & \left.
-\langle [B_1 B_2]^\CP_{II_z} | G(\bp, -\bp^\prime; K, \omega)P_\sigma
P_F | [B_3 B_4]^{\CP^\prime}_{II_z}\rangle \right\}\ \ .
\label{a8}
\end{eqnarray}
Here the second term in the brackets gives the same contribution
as the first term, due to the condition \eq{a5}.
Thus we find that \eq{a7} is actually invariant $G$-matrix,
which allows the expressions
\begin{eqnarray}
& & G_{ca}(\bp, \bp^\prime; K, \omega) \nonumber \\
& & =\langle [B_1 B_2]_{II_z} | G(\bp, \bp^\prime; K, \omega)
-G(\bp, -\bp^\prime; K, \omega) P_\sigma P_F | [B_3 B_4]_{II_z} \rangle
\nonumber \\
& & =\sqrt{(1+\delta_{c_1,c_2})(1+\delta_{a_1,a_2})}
\sum_{\CP, \CP^\prime}
\langle [B_1 B_2]^\CP_{II_z} | G(\bp, \bp^\prime; K, \omega) |
[B_3 B_4]^{\CP^\prime}_{II_z}\rangle
\nonumber \\
& & =\sqrt{(1+\delta_{c_1,c_2})(1+\delta_{a_1,a_2})}
\sum^\prime_{JM\ell \ell^\prime S S^\prime \CP \CP^\prime}
4\pi\,G^J_{\gamma S\ell,\alpha S^\prime{S_z}^\prime}(p, p^\prime;
K, \omega)
\nonumber \\
& & \qquad \times \CY_{(\ell S)JM}(\widehat{\bp}; {\rm spin})
\,\CY^*_{(\ell^\prime S^\prime)JM}(\widehat{\bp}^\prime; {\rm spin})
\nonumber \\
& & =\sqrt{(1+\delta_{c_1,c_2})(1+\delta_{a_1,a_2})}
\left[ g_0 + g_{ss}\,(\bfsigma_1\cdot \bfsigma_2)
+h_0\,i\widehat{\bn}\cdot (\bfsigma_1+\bfsigma_2)
\right. \nonumber \\
& & \left. \qquad +h_-\,i\widehat{\bn}\cdot (\bfsigma_1-\bfsigma_2)
+\cdots \right]\ \ .
\label{a9}
\end{eqnarray}
The eight independent invariant functions, $g_0$, $g_{ss}$, etc.,
are expressed by some combinations of the
partial-wave components of the $G$-matrix, \linebreak
$G^J_{\gamma S\ell,\alpha S^\prime{S_z}^\prime}(p, p^\prime;
K, \omega)$, which are explicitly given in Appendix D of
Ref.\,\cite{LSRGM}.

\end{document}